\documentclass[twocolumn]{aastex63}

\usepackage{multirow}
\usepackage{tabularx}
\usepackage{graphicx}
\usepackage{float}
\usepackage{epstopdf}
\usepackage{subfigure}
\usepackage{enumerate}
\usepackage{threeparttable}
\usepackage{rotating}
\usepackage{amsmath}
\usepackage{diagbox}

\allowdisplaybreaks

\begin{document}

\title{WIYN $^5$ Open Cluster Study LXXX: HDI CCD $UBVRI$ Photometry of the Old Open Cluster NGC 7142 and Comparison to M67}

\author{Qinghui Sun}
\author{Constantine P. Deliyannis $^4$}
\affil{Department of Astronomy, Indiana University, Bloomington, IN 47405, USA \\ qingsun@indiana.edu; cdeliyan@indiana.edu}
\author{Bruce A. Twarog}
\author{Barbara J. Anthony-Twarog}
\affil{Department of Physics and Astronomy, University of Kansas, Lawrence, KS 660045, USA \\ btwarog@ku.edu; bjat@ku.edu}
\author{Aaron Steinhauer}
\affil{Department of Physics and Astronomy, State University of New York, Geneso, NY 14454, USA \\ steinhau@geneseo.edu}

\footnote[4]{Visiting Astronomer, Kitt Peak National Observatory, National Optical Astronomy Observatory, which is operated by the Association of Universities for Research in Astronomy (AURA) under cooperative agreement with the National Science Foundation.}
\footnote[5]{The WIYN Observatory is a joint facility of the University of Wisconsin-Madison, Indiana University, the National Optical Astronomy Observatory and the University of Missouri.}

\begin{abstract}

We present $UBVRI$ photometry of 8702 stars in a $0.5 ^\circ\times0.5 ^\circ$ field in the direction of NGC 7142, taken with the Half Degree Imager (HDI) at the WIYN 0.9-meter telescope, to improve knowledge of this cluster's basic parameters. Our photometry spans the ranges 10.6 - 20.4 mag in $U$, 10.6 - 22.0 mag in $B$, 10.0 - 21.8 mag in $V$, 9.2 - 20.7 mag in $R$, and 8.5 - 19.9 mag in $I$. Using color-color diagrams that employ all four color combinations that include $U$, versus $B-V$, we derive a reddening-metallicity relation for the cluster, with preferred values $E(B-V$) = 0.338 $\pm$ 0.031 mag for the left-edge fiducial of the main sequence and [Fe/H] = 0.0 $\pm$ 0.1 dex, where the Hyades cluster has been used as an unreddened reference cluster, the extinction relations of Cardelli have been employed, and the metallicity dependence of the Yonsei-Yale ($Y^2$) isochrones has been assumed. Comparison of our data to the $Y^2$ isochrones in multiple color-magnitude diagrams (CMDs) yields distance-metallicity and age-metallicity relations, with preferred values of $m-M\ =\ 12.65\ \pm\ 0.23$ mag and age = $4.0 ^{-0.7}_{+1.3}$ Gyr. Re-evaluation of the parameters of M67 using Stetson's $UBVI$ photometry yields [Fe/H] = -0.02 $\pm$ 0.05 dex, $E(B-V$) = 0.04 $\pm$ 0.01 mag, $m-M$ = 9.75 $\pm$ 0.03 mag, and age = 3.85 $\pm$ 0.17 Gyr; we thus find the metallicity and age of the two clusters to be indistinguishable. A semi-independent analysis adopting the parameters of M67 and shifting the fiducials of the two clusters in six CMDs until they match strongly corroborates the values listed above. The differences between our inferred parallaxes and the Gaia DR2 values are $87\ \pm\ 60$ $\mu$as for NGC 7142 and 48 $\pm$ 15 $\mu$as for M67, consistent with previous studies.

\end{abstract}


\section{Introduction}  \label{Introduction}

Old open clusters are important probes of cluster and Galactic evolution, especially the structure, kinematics, and chemical evolution of the early Galactic disk \citep{Friel95,Carraro13}. Their old stars also probe stellar evolution, e.g., in helping to ascertain the size of the mixed core \citep{Dinescu95} or to clarify the action of physical mechanisms that affect abundances of light element probes such as Li \citep{Deli94}. However, old open clusters are relatively rare and often hard to identify and separate from the rich background Galactic field; their ages, distances, chemical compositions, and reddenings can also be challenging to determine with confidence \citep{JH11}.

NGC 7142, located at the Galactic ({\it l, b}) = ($105.347\arcdeg$, $+9.485\arcdeg$), is a potentially important old open cluster. Various photometric and spectroscopic studies have been performed to determine cluster parameters such as age, metallicity, distance, and chemical abundances. Photometric studies began to appear in the 1960s and 1970s, including $UBV$ photoelectric (pe) and/or photographic (pg) measurements \citep[hereafter S69, H61, VH70, and JH75; JH75 also included $iyz$ and S69 presented only pg $BV$]{Hoag61,Sharov69,VH70,JH75}. Using the photometry of H61, \citet[B62]{Bergh62} was the first to suggest that the color-magnitude diagram (CMD) of NGC 7142 bore similarities to those of the already famous old open clusters M67 and NGC 188. The existence of variable reddening across the face of the cluster was suggested by \citet[S65]{Sharov65} on the basis of star counts, and by VH70 from the appearance of the region in the Palomar Observatory Sky Survey. \citet[C86]{Canterna86} derived a cluster metallicity approximately 0.1 dex lower than that of the Hyades using photoelectric Washington photometry. In a search for variable stars, \citet[CT91]{CT91} published the first CCD photometry of the cluster, using the $BV$ bands. CT91 concurred about the existence of variable reddening, and suggested that its extent is less than 0.1 mag in $E(B-V$). NGC 7142 did not receive much attention again until the 2010s. \citet[JH11]{JH11} published a CCD $BVI$ study, while \citet[S11]{Sandquist11} published a CCD $BVRI$ study. The latter reported detection of eight variables, most of which were believed to be cluster members. Performing an ultraviolet-oriented imaging survey, \citet{Carraro13} reported CCD $UB$ photometry. Finally, \citet[S14]{Straizys14} published Vilnius photometry and suggested that the range of extinction across the cluster may be somewhat larger than indicated above. A medium-resolution spectroscopic study \citep[FJ93]{Friel93} and two high-resolution studies \citep[J07, J08]{Jacobson07, Jacobson08} have also been conducted. Results from these studies often conflict, and there remain significant uncertainties about the basic cluster parameters.

Regarding the cluster reddening, using the H61 data, \citet[J61]{Johnson61} and \citet[BF71]{Becker71} determined $E(B-V$) = 0.18 mag and 0.19 mag, respectively. However, while VH70 also used pe/pg data and employed similar techniques (color-color diagrams), perhaps surprisingly they derive the much higher value of $E(B-V$) = 0.41 mag. While VH70 studied more stars than did H61, and in particular, included more photoelectric observations, VH70 barely reached as faint as the cluster turnoff (TO). S11 point out that even higher reddening values, of order 0.51, are possible, as inferred from IRAS \footnote{The Infrared Astronomical Satellite} and COBE \footnote{The Cosmic Background Explorer} dust maps, although it is possible that this value applies beyond the cluster. Using only five giants, JH75 derived an intermediate value of $E(B-V$) = 0.29 mag, and a reanalysis of what CT91 considered a more appropriate subset of just two of the JH75 giants yielded $E(B-V$) = 0.35 mag. JH11 found $E(B-V$) = 0.32 $\pm$ 0.05 mag by comparing their CCD $BVI$ data for the red clump and TO to synthetic CMDs from Padova isochrones \citep{Girardi00, Marigo08}. Consistent with this, to within the rather large stated uncertainties, S11 found $E(B-V$) = 0.25 $\pm$ 0.06 mag from optical and Two Micron All Sky Survey (2MASS) $JHK$ comparisons of clump stars in NGC 7142 and in M67, using the method of \citet{Gro02}, and assuming reasonable values for the parameters of M67. However, S11 also derived their preferred value of $E(B-V$) = 0.25 $\pm$ 0.06 mag, using the method of \citet{McCall04}. S14 determined an average cluster $E(B-V)\ \sim$ 0.35 mag from their Vilnius photometry. Note that none of these studies except S14 employed main sequence (MS) stars that are located clearly below the MS TO, and neither J11 nor S11 used the $U$ filter, which is very sensitive to reddening and metallicity.

The cluster metallicity is also uncertain. Based on their photometry, JH75 derived [Fe/H] = -0.45 $\pm$ 0.2 dex, and (using eight stars) C86 derived [Fe/H] = -0.1 dex on a scale where the Hyades is [Fe/H] = 0.0 $\pm$ 0.15 dex. From their medium-resolution spectra, FJ93 derived [Fe/H] = -0.23 $\pm$ 0.13 dex, which was revised upward to -0.10 $\pm$ 0.10 dex by \citet{Friel02} from a new calibration, and to +0.04 $\pm$ 0.06 dex by \citet{Twarog97} by transforming the same data to a common metallicity scale based on DDO photometry. From two different sets of high-resolution spectroscopic data for the same four giants, J07 report [Fe/H] = +0.08 $\pm$ 0.06 dex and J08 report [Fe/H] = +0.14 $\pm$ 0.01 dex.

Regarding the distance modulus, reported values from the above studies range in $(m-M)_V$ from 11.8 mag to 13.7 mag and $(m-M)_0$ from 10.9 mag to 12.5 mag. VH70 find $(m-M)_0$ = 12.5 mag, JH75 find 10.9 mag, CT91 find 11.4 mag, and J11 find 11.85 $\pm$ 0.05 mag. In good agreement with J11, S11 find 11.96 $\pm$ 0.15 mag. S14 report $\sim$ 11.8 mag using their and 2MASS photometry of five red clump stars identified in S11.

Reported or inferred ages cover a disappointingly large range, from about 2 Gyr to about 7 Gyr. VH70 concluded that NGC 7142 is older than NGC 7789 and younger than M67; modern estimates for the TO ages of these clusters include 1.5 $\pm$ 0.1 Gyr for NGC 7789 \citep{Brunker13} using $Y^2$ isochrones \citep{Kim02, Demarque04} and 4.0 $\pm$ 0.5 Gyr for M67 using precursor models to the $Y^2$ isochrones (\citet[S99]{Sarajedini99}; \citet{Sills00}). CT91 concluded that NGC 7142 is older than M67 but younger than NGC 188 (7 $\pm$ 0.5 Gyr old; S99). \citet{Carraro98} report an age of 4.9 Gyr using synthetic CMDs, while \citet{Salaris04} report an age of 4 $\pm$ 1 Gyr from the difference in brightness between the red clump and the subgiant branch. By comparing to CMDs of M67, S11 conclude NGC 7142 is slightly younger than M67, around 3 Gyr old. S14 also report 3.0 $\pm$ 0.5 Gyr. However, whereas most of the above estimates are close to the age of M67, JH11 find an age of 6.9 Gyr, much closer to that of NGC 188, based on synthetic CMDs.

Improving knowledge of the age of NGC 7142 is particularly important for studies that use Li as a probe of stellar interiors. The deepening surface convection zones of subgiants evolving out of M67 reveal the physical mechanisms occurring during the MS that affect the surface Li abundance \citep{Sills00}. The M67 subgiants evolve out of the deep part of the cool side of the ``Li Dip" phenomenon \citep{Boesgaard86, Cummings17}. Subgiants in NGC 188 evolve out of the shallower part of cool side of the Li Dip, providing information complementary to that of M67. A cluster whose subgiants are evolving out of the hot side of the Li Dip, and thus slightly younger than M67, would provide invaluable information about physical mechanisms involved in forming the hot side of the Li Dip. In principle, these mechanisms could be different than those responsible for the cool side. Furthermore, there are few accessible clusters of appropriate age. Clarification of the age of NGC 7142 could thus be of great importance to such studies.

The goal of this study is to refine our knowledge of the basic parameters of NGC 7142, using broadband $UBVRI$ photometry. We mainly use stars well below the MS TO, which are particularly advantageous as tools to determine the basic cluster parameters. We also must stress that we employ the $U$ filter, which is very sensitive to both reddening and metallicity.

The paper is organized as follows. Section \ref{obs and redu} describes our observations, data processing, and reduction procedures, and compares our photometry to those from previous studies; Section \ref{analysis} discusses the cluster's CMDs, estimates the cluster's metallicity and reddening from color-color diagrams, and derives the cluster's distance and age; and Section \ref{summary} summarizes our findings.

\section{Observations and Data Reductions} \label{obs and redu}

\subsection{Observations and Processing}

\begin{figure*}
	\centering
	\includegraphics[width=0.8\textwidth]{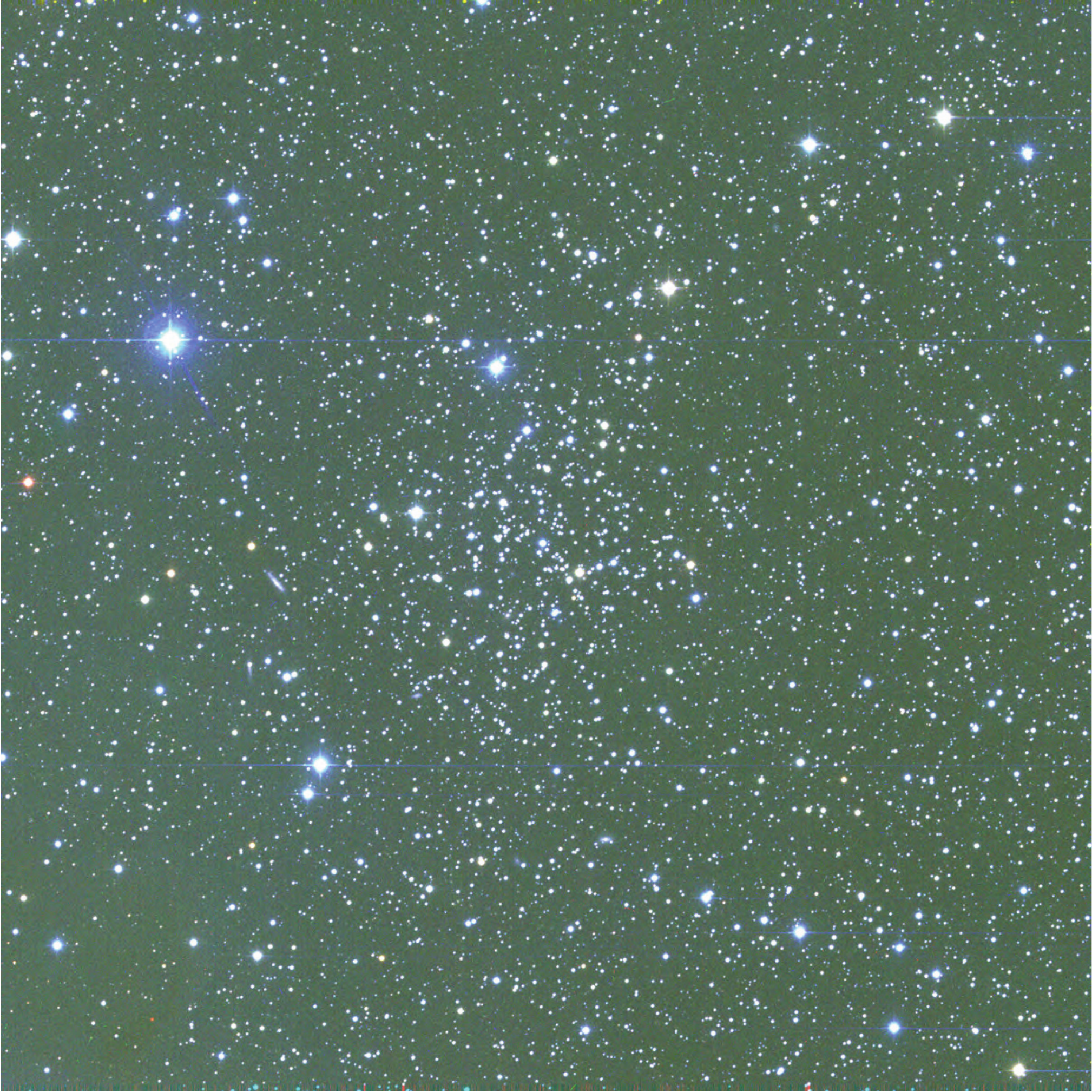} 
	\caption{RGB image of NGC 7142, the image covers a $0.5 ^\circ \times0.5 ^\circ$
		field of view at 0.43$\arcsec$ pixel$^{-1}$}
	\label{fig:sample image}
\end{figure*}

$UBVRI$ images of NGC 7142 were taken under photometric conditions using the WIYN 0.9m telescope at the Kitt Peak National Observatory on the night of 2015 December 5, with the Half Degree Imager (HDI) using an e2V 231-84 $4096\times4096$ CCD detector, covering a $0.5^\circ \times 0.5^\circ$ field (at 0.43$\arcsec$ pixel$^{-1}$) around the cluster center. Table \ref{tab:observing log} shows a log of the observations, and Figure \ref{fig:sample image} shows a color image of the field of view made from our data. Dithering was employed to place stars on different pixels with each pass of $UBVRI$ exposures. For each filter, extra-short (1s), short (5s), medium (25s), long (125s), and extra-long (625s) exposures were taken, providing a large dynamic range. For example, we report photometry for stars ranging over 10.5 mag in $V$, from $V$ = 10.1 to 21.7 mag. Processing (debiasing, flat-fielding, and illumination correction) of our CCD images was carried out using the {\it ccdproc} routine in IRAF\footnote{IRAF is distributed by the National Optical Astronomy Observatories, which are operated by the Association of Universities for Research in Astronomy Inc., under cooperative agreement with the National Science Foundation.}. Additional details, as well as more detailed discussions of our reduction procedures for standard and object images can be found in \citet[in preparation]{Deli20}.

\begin{table}[htbp!]
	\caption{Photometric Observations of NGC 7142}
	\label{tab:observing log}
	\centering	
	\begin{tabular}{|c|c|c|c|c|}
		\hline
		Date & Exposure (s) & Filter & Airmass \\
		\hline
		 & {In each filter: $4\times1$ s} & $U$ & $1.293 -1.656$\\ 
		{2015 Dec 5} & $3\times5$ s & $B$ &  $1.271 - 1.666$\\ 
		2015 & $3\times25$ s & $V$ &  $1.271 - 1.677$\\ 
		& $3\times125$ s  & $R$ &  $1.270 - 1.688$\\ 
		& $1\times625$ s & $I$ &  $1.269 - 1.699$\\ 
		\hline
	\end{tabular}
\end{table}

\subsection{Analysis of Standard Stars} \label{standards}
 
Standard stars were drawn from the catalogs of \citet{Landolt92, Landolt09}. An aperture size of 40 pixels in diameter ($\sim$ 17 $\arcsec$) was used to measure the instrumental magnitudes of the standard stars. This aperture is just slightly larger than most apertures used by Landolt, and we verified that it contains $\gtrsim 99\%$ of the light of representative stars. We used the following transformation equations:

\begin{eqnarray}
u &=& U + a_U +b_U^1(B-V)+b_U^2(B-V)^2+ b_U^3(B-V)^3 \nonumber  \\
  & & + b_U^4(B-V)^4+ b_U^5(B-V)^5+c_UX_U \label{eqn1}\\
b &=& B + a_B +b_B(B-V)+c_BX_B \label{eqn2} \\
v &=& V + a_V +b_V(B-V)+c_VX_V \label{eqn3} \\ 
v &=& V + a_V +b_V'(V-I)+c_V'X_V \label{eqn4} \\
r &=& R + a_R +b_R(V-R)+c_RX_R \label{eqn5} \\
i &=& I + a_I +b_I(V-I)+c_IX_I \label{eqn6}
\end{eqnarray}

\noindent where the uppercase letters are Landolt magnitudes, lowercase letters $ubvri$ are instrumental magnitudes, $X$ is the airmass, and lowercase letters $a, b,$ and $c$ are the zero points, color-term coefficients, and extinction coefficients, respectively, the values of which were derived from fits of the above equations to our measurements. The fits were carried out using IRAF's {\it fitparams} task in the {\it plotcal} package. Neither the cross terms of airmass and color nor the UT terms are significant, so none of them are shown above. Table \ref{tab:fit standards} shows the coefficient values and errors from our best (and adopted) fits. The coefficient values are reasonable: for example, the extinction coefficients are on the low side of the known range of values for photometric nights at Kitt Peak.

\begin{table*} 
	\renewcommand{\arraystretch}{1.3}
	\caption{Coefficients for $UBVRI$ calibration and valid color range}
	\centering
	\footnotesize
	\begin{tabular}{|c|c|c|c|c|c|c|}
		\hline
		Filter & \# of stars & RMS & $a$ & $b$ & $c$ & Color range (mag) \\
		\hline
		$U$ & 160 & 0.026 & 3.673 $\pm$ 0.011 & $b^1=-0.686 \pm 0.028$ &
		 0.4036 $\pm$ 0.0063 & --0.2 - +2.1\\
		 & & & & $b^2 = 2.604 \pm 0.1298$ & & \\
		 & & & & $b^3 = -3.163 \pm 0.2249$ & & \\
		 & & & & $b^4 = 1.505 \pm 0.1549$ & & \\
		 & & & & $b^5 = 0.250 \pm 0.0362$ & & \\
		\hline
		$B$ & 162 & 0.015 & 1.768 $\pm$ 0.006 & -0.037 $\pm$ 0.0020 & 0.2260 $\pm$ 0.0037 & --0.3 - +2.1\\
		\hline
		$V_{B-V}$ & 172 & 0.012 & 1.820 $\pm$ 0.005 & 0.025 $\pm$ 0.0015 & 0.1390 $\pm$ 0.0028 & --0.3 - +2.2\\
		\hline
		$V_{V-I}$ & 173 & 0.012 & 1.820 $\pm$ 0.005 & 0.024 $\pm$ 0.0014 & 0.1388 $\pm$ 0.0028 & --0.3 - +2.5\\
		\hline
		$R$ & 154 & 0.020 & 1.765 $\pm$ 0.008 & -0.008 $\pm$ 0.0048 & 0.0859 $\pm$ 0.0049 & --0.2 - +1.3\\
		\hline
		$I$ & 175 & 0.018 & 2.328 $\pm$ 0.007 & -0.050 $\pm$ 0.0020 & 0.0438 $\pm$ 0.0040 & --0.4 - +2.5\\
		\hline
	\end{tabular}
	\label{tab:fit standards}
\end{table*}

\begin{figure}
	\centering
		\includegraphics[width=0.45\textwidth]{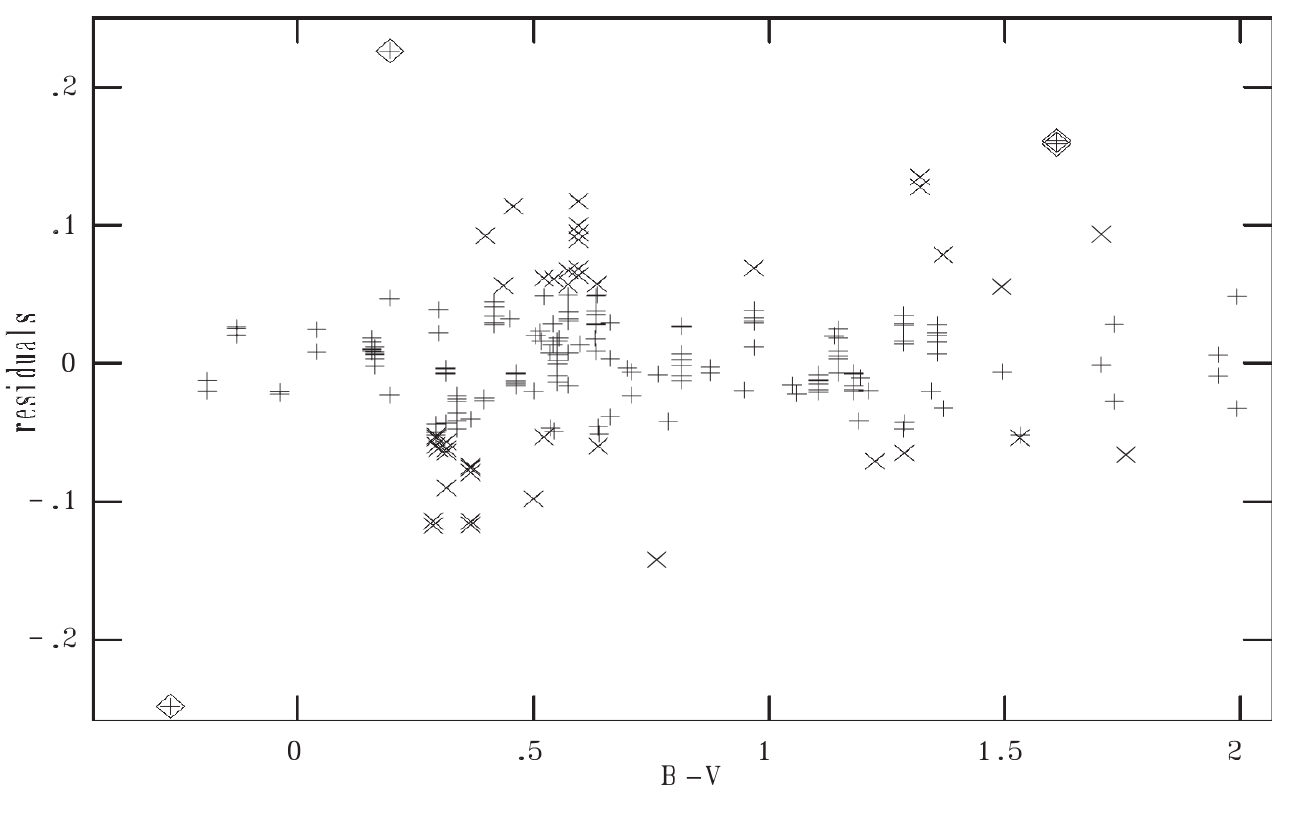}
	\caption{Residuals of the fit to the standard star $U$ transformation (equation (\ref{eqn1}) versus $\bv$. Stars retained in the fit appear as plain + signs. Rejected stars appear as X signs or + signs within squares.} 
	\label{fig:res vs BV}
\end{figure}

\begin{figure}
	\centering
	\includegraphics[width=0.45\textwidth]{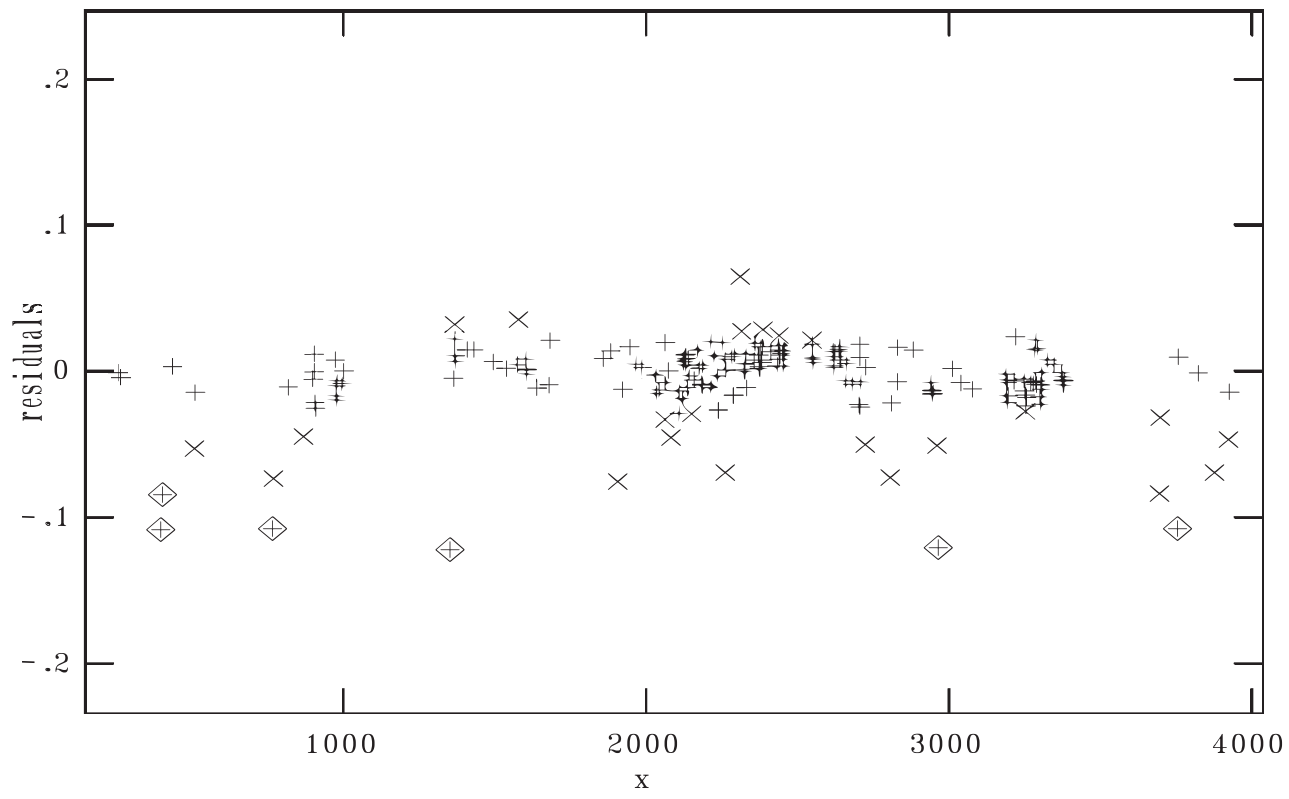}
	\caption{Residuals of the fit to the standard star $V$ transformation (equation (\ref{eqn3}) versus x. Same symbols as in Figure \ref{fig:res vs BV}.}
	\label{fig:res vs x}
\end{figure}

In performing the fits, outliers were rejected by IRAF and by us in the residuals versus fitting function plane only. While linear color terms worked well for $BVRI$, having also explicitly verified that quadratic color terms are insignificant, a higher-order polynomial in color is needed when calibrating the $U$ filter. We use the $B-V$ color for the $U$ calibration. As with the (previous) S2KB imager of the WIYN 0.9m \citep[in preparation]{Deli20}, a distinct improvement to the fit was found when going from a cubic to a quartic polynomial. Addition of a fifth-order term was not only statistically significant (Table \ref{tab:fit standards}) but also resulted in a flatter relation between the residuals to the fit and the ($B-V$) color (Figure \ref{fig:res vs BV}). We also tried a sextic polynomial, but no further improvements resulted, so we adopted the quintic polynomial for $U$. None of the residuals for any of the filters showed any spatial dependence in either the $x$ or $y$ directions (Figure \ref{fig:res vs x}, residuals for the $V$ calibration versus $x$). The absence of any UT dependence for all filters corroborates that the night was photometric. Column 3 of Table \ref{tab:fit standards} shows the RMS of the residuals, and Column 7 shows the color range covered by the standards (using the colors indicated in the transformation equations for each filter), and thus the color range over which our transformations are valid. We caution the reader not to use photometry for stars with colors outside of these ranges, {\it especially for the $U$ filter}.

\subsection{Reductions of NGC 7142 images} \label{reduction}
 
\begin{figure}
	\centering
	\includegraphics[trim={0cm 0.5cm 1cm 0.2cm},clip,width=0.45\textwidth]{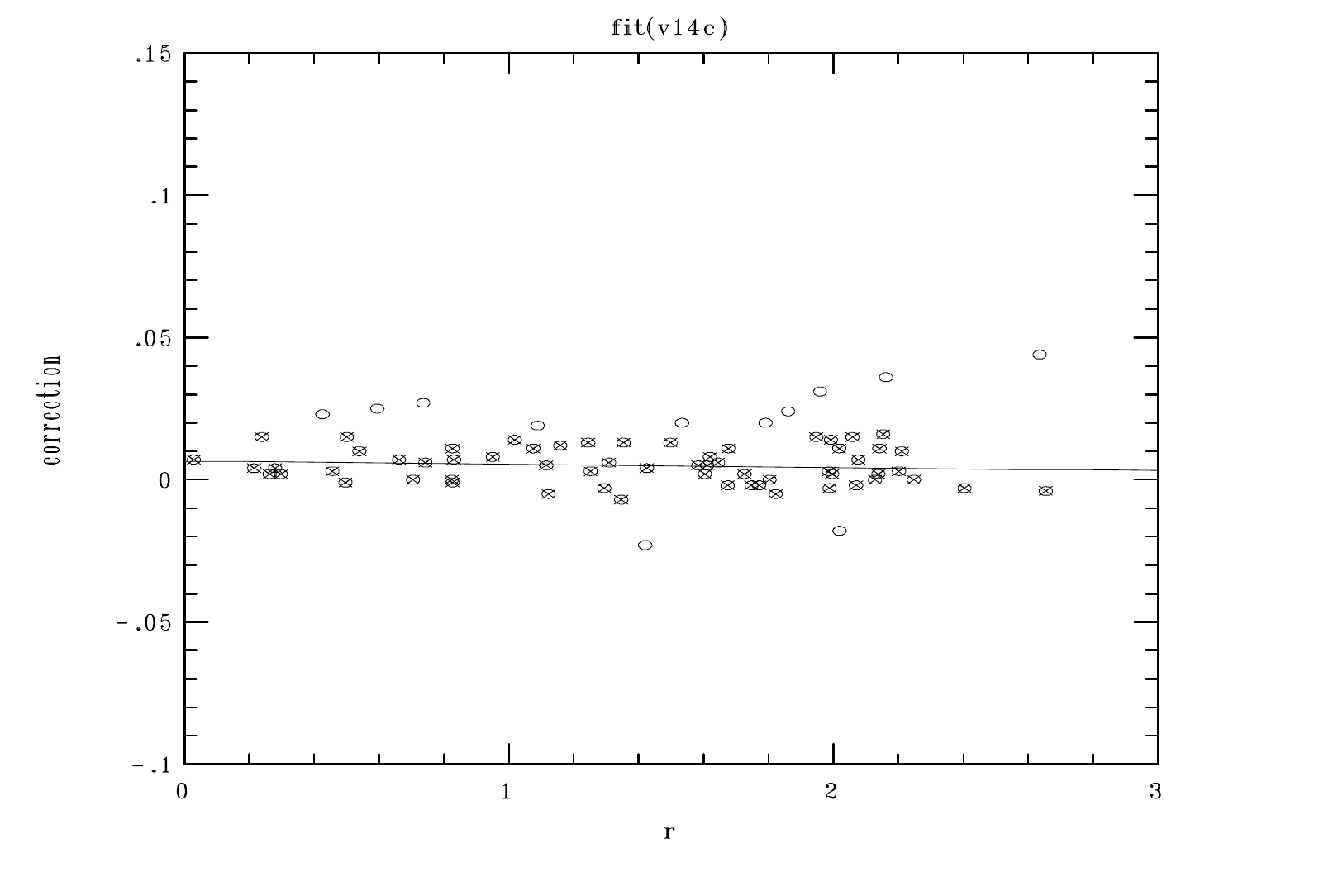} 
	\caption{Aperture correction for a sample $U$ (filter image), as a function of r (pixels) from the image center}
	\label{fig:aper vs r}
\end{figure}

We employed IRAF, using the standalone ALLSTAR and DAOPHOT III \citep{Stetson87} software packages to extract stellar photometry from NGC 7142 images. For each image, a high signal-to-noise point spread function (PSF) was constructed using between 80 and 300 bright ( $<$ 13 mag), unsaturated, linear, isolated stars, and the PSF was allowed to vary up to quadratically across the image. For some of the shorter $U$ and shortest $BVRI$ exposures, the typical number of bright stars was only between 40 and 80, and a spatially constant PSF was constructed. After the PSF was applied to all stars in the image, stars with errors $>$ 0.1 mag were discarded, as were those bright stars (if any) with evidence for nonlinearity. Further edits were applied based on sharpness, CHI value, and excess scatter from the mean trend. We applied aperture corrections to the PSF-based magnitudes to get the total stellar magnitudes. Based on bright, isolated stars, most of the frames have constant aperture corrections with respect to spatial variables $x$, $y$, and $r$, and only a few have slightly linear slopes; typical aperture corrections are 0.003 - 0.013 mag; an example is shown in Figure \ref{fig:aper vs r}. 

\subsection{Photometric calibration of cluster images}
 
\begin{figure}
	\centering
	\includegraphics[width=0.55\textwidth]{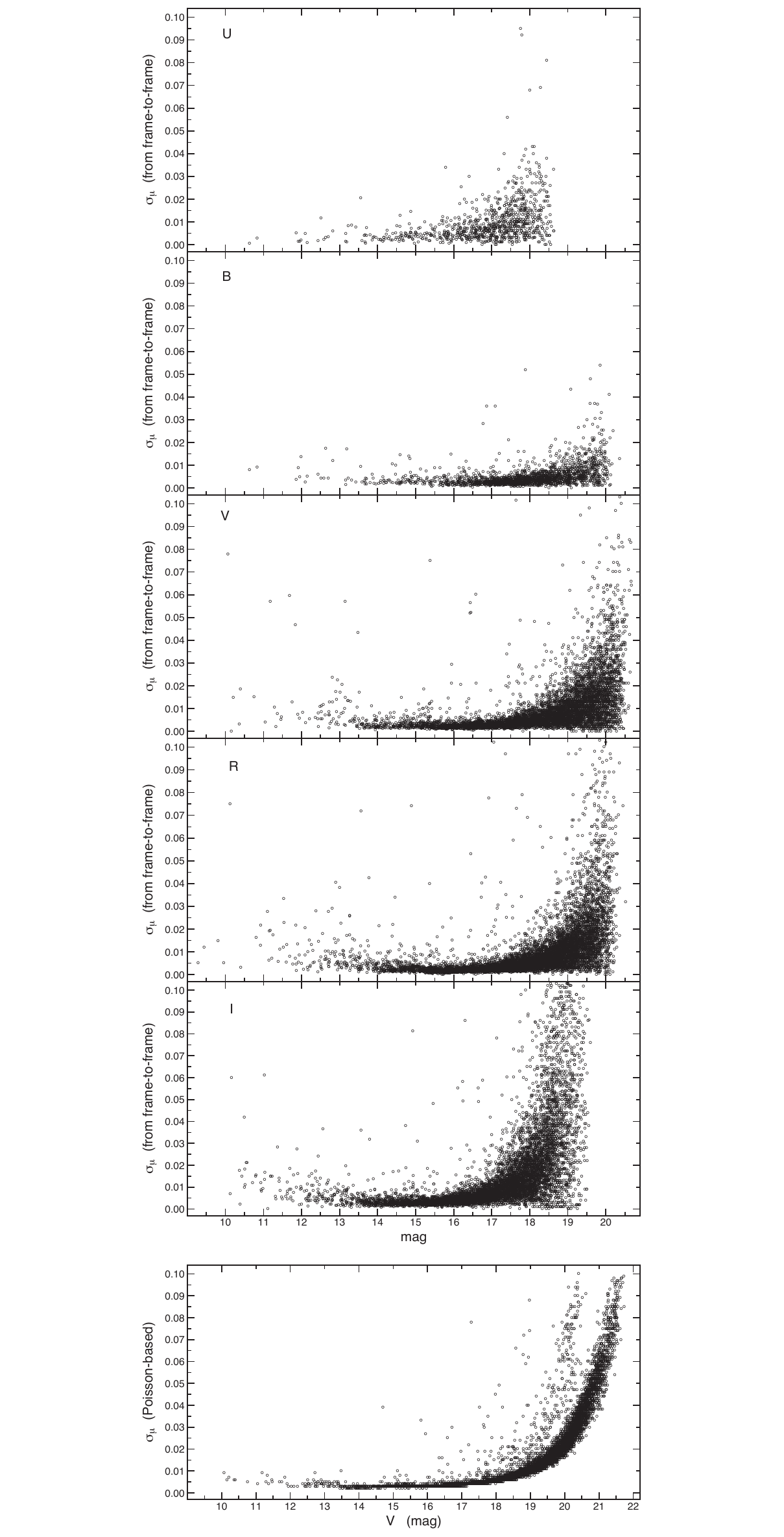} 
	\caption{Errors as a function of magnitude.}
	\label{fig:error vs mag}
\end{figure}

The images were matched and combined using DAOMATCH and DAOMASTER, and transformed using equations (\ref{eqn1}) - (\ref{eqn6}). For each star, we report photometry in at least two filters, which always include the $V$ magnitude. The $V$ magnitude was calibrated using the $B-V$ color (equation (\ref{eqn3})), except for stars too faint to have a detected $B$ magnitude. In those cases where $V$ and $I$ are both detected, $V$ was calibrated using the $V-I$ color (equation (\ref{eqn4})). We report photometry for 8402 stars, including $UBV$ for the brightest stars in the field. We present our final photometry in Table \ref{tab:photometry}, which appears entirely online. Columns 1-5 are WOCS ID (see \citet[in preparation]{Deli20}; a cluster center at R.A. = $21^h45^m10.29^s$, decl. = $65^\circ46\arcmin38.4\arcsec$ was assumed); $x$, $y$ positions in pixels; and the corresponding R.A., and decl. (equinox 2000) determined with the aid of the USNO catalog,\footnote{http://vizier.u-strasbg.fr/viz-bin/VizieR} with typical expected errors in each variable of 0.153$\arcsec$. For each filter, we report the calibrated magnitude, Poisson-based $\sigma_{\mu}$ as combined in DAOMASTER from the Poisson-based $\sigma$ in each frame as reported by DAOPHOT, and frame-to-frame-based sigma from DAOMASTER. Finally, the number of observations is indicated along with the 10 colors one can construct from these five filters. No measurement is indicated by ``99.999" or ``9.999." For each filter, Figure \ref{fig:error vs mag} shows the final $\sigma_{\mu}$ based on the frame-to-frame dispersion reported by DAOMASTER, as a function of magnitude; for comparison, the purely Poisson-based $\sigma_{\mu}$ is shown for $V$. Note that faint stars (e.g. $V\ \geq 20.5$ mag) that are detected only in the single long (625 s) exposure appear only in the bottom panel and not in the upper panel, since $\sigma_{\mu}$(Poisson) requires only one measurement whereas $\sigma_{\mu}$(dispersion) requires at least two.

\begin{deluxetable*}{rccccccc}
	\tablewidth{0pt}
	\tablecaption{Photometry for stars in NGC 7142}
	\label{tab:photometry}
	\tablehead{
		\colhead{ID} & \colhead{R.A. (J2000)} & \colhead{decl. (J2000)} &
		\colhead{U} & \colhead{B} & \colhead{V} & \colhead{R} & \colhead{I}
	}
	\startdata
	1017&21 44 13.55&65 52 38.2&10.830&10.680&10.064&9.703&99.999 \\
	1010&21 45 56.15&65 47 47.2&10.632&10.642&10.154&99.999&9.633 \\
	1732&21 46 58.15&65 35 45.5&13.682&11.830&10.206&9.277&99.999 \\
	\enddata
	\tablecomments{The full version contains $X$ and $Y$ pixel coordinates, uncertainties on the photometry, number of observations in each $UBVRI$ filter, and colors. (This table is available in its entirety in machine-readable form.)}
\end{deluxetable*}

\subsection{Comparisons to previous NGC 7142 photometry} \label{sec:compare}

\begin{figure}
	\centering
    \includegraphics[trim={0.2cm 2.5cm 1cm 0.2cm},clip,width=0.5\textwidth]{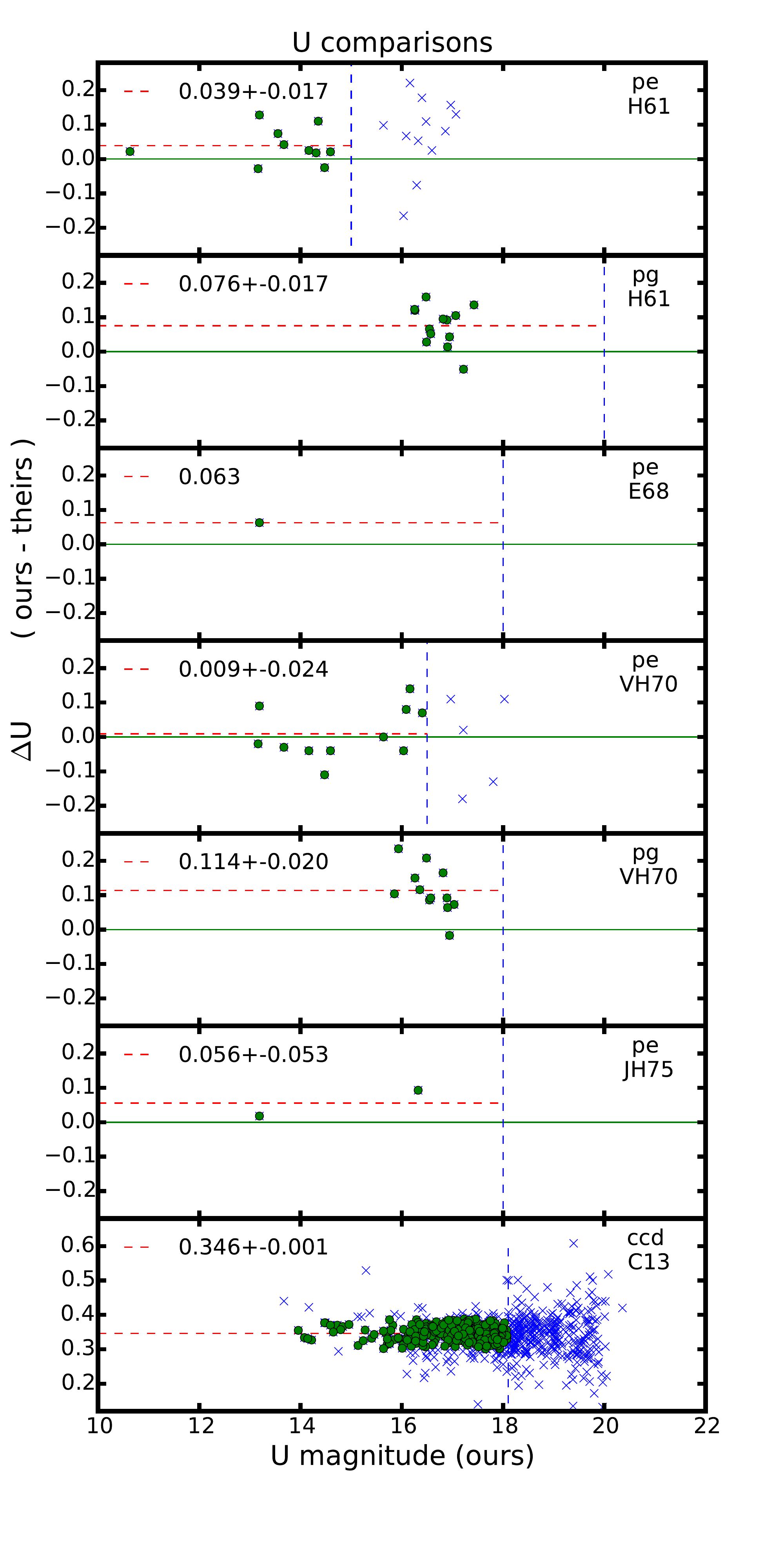}\\
    \caption{Comparison to previous studies in $U$ filters}
    \label{fig:compare U}
\end{figure}

\begin{figure}
	\centering
	\includegraphics[trim={0.2cm 4.5cm 1cm 0.2cm},clip,width=0.45\textwidth]{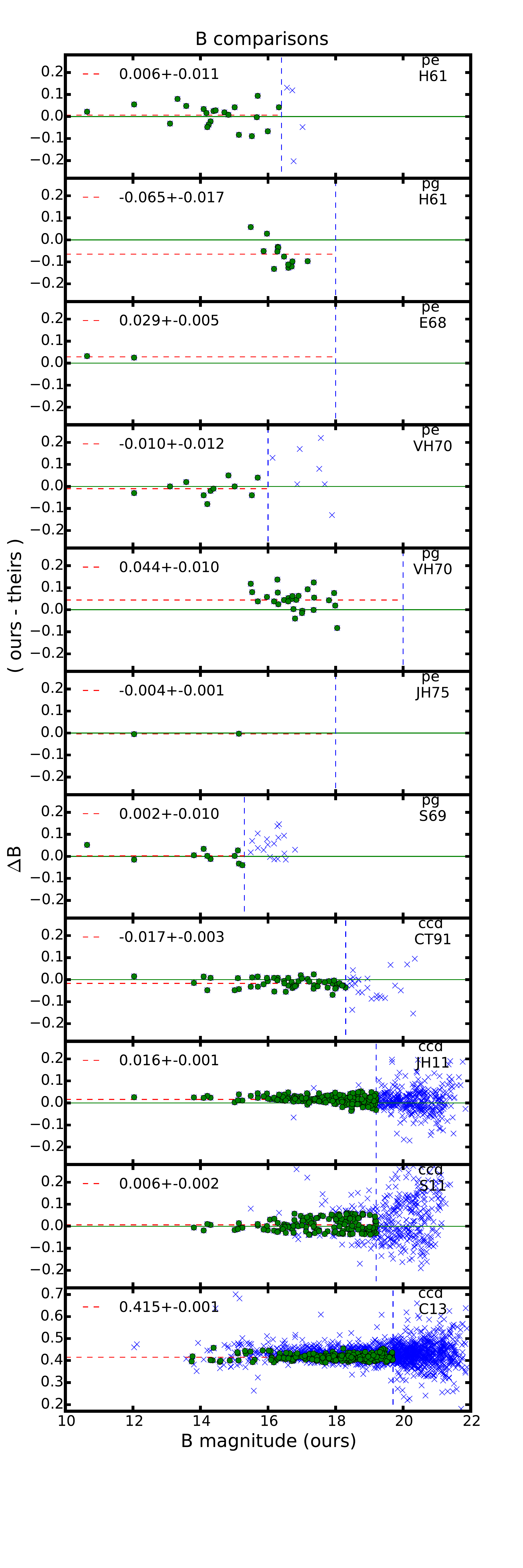}\\
	\caption{Comparison to previous studies in $B$ filters}
	\label{fig:compare B}
\end{figure}

\begin{figure}
	\centering
	\includegraphics[trim={0.2cm 4.5cm 1cm 0.2cm},clip,width=0.45\textwidth]{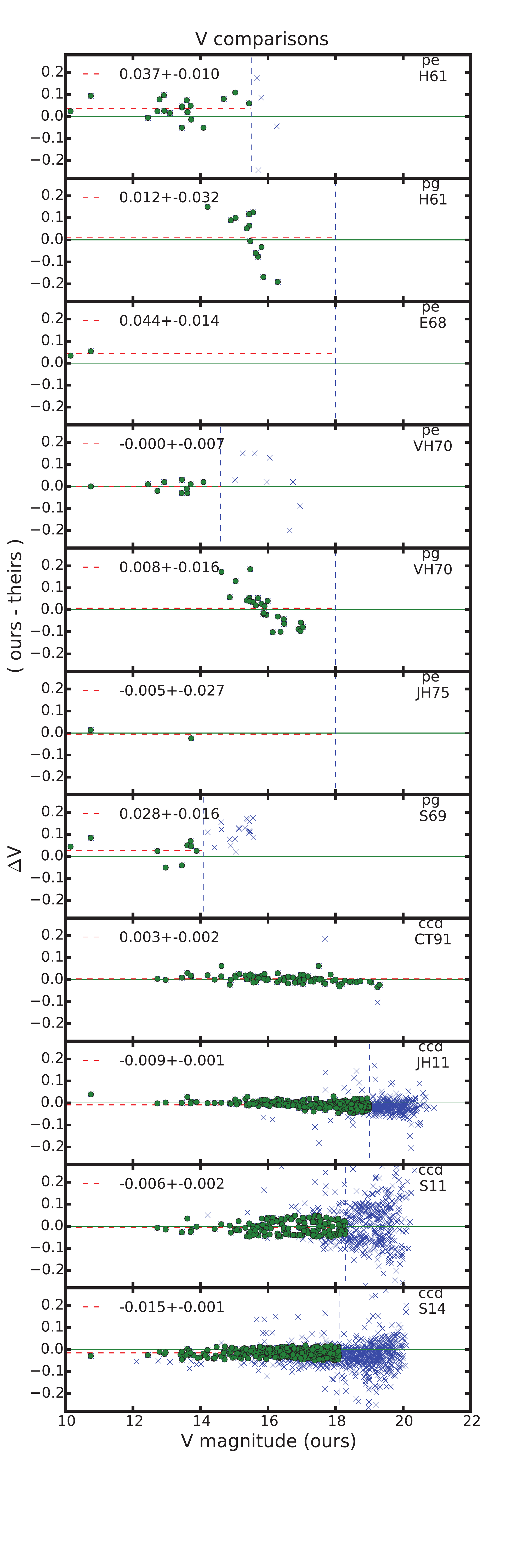}\\
	\caption{Comparison to previous studies in $V$ filters}
	\label{fig:compare V}
\end{figure}

\begin{figure}
	\includegraphics[trim={0.20cm 0.cm 0.4cm 0.2cm},clip,width=0.5\textwidth]{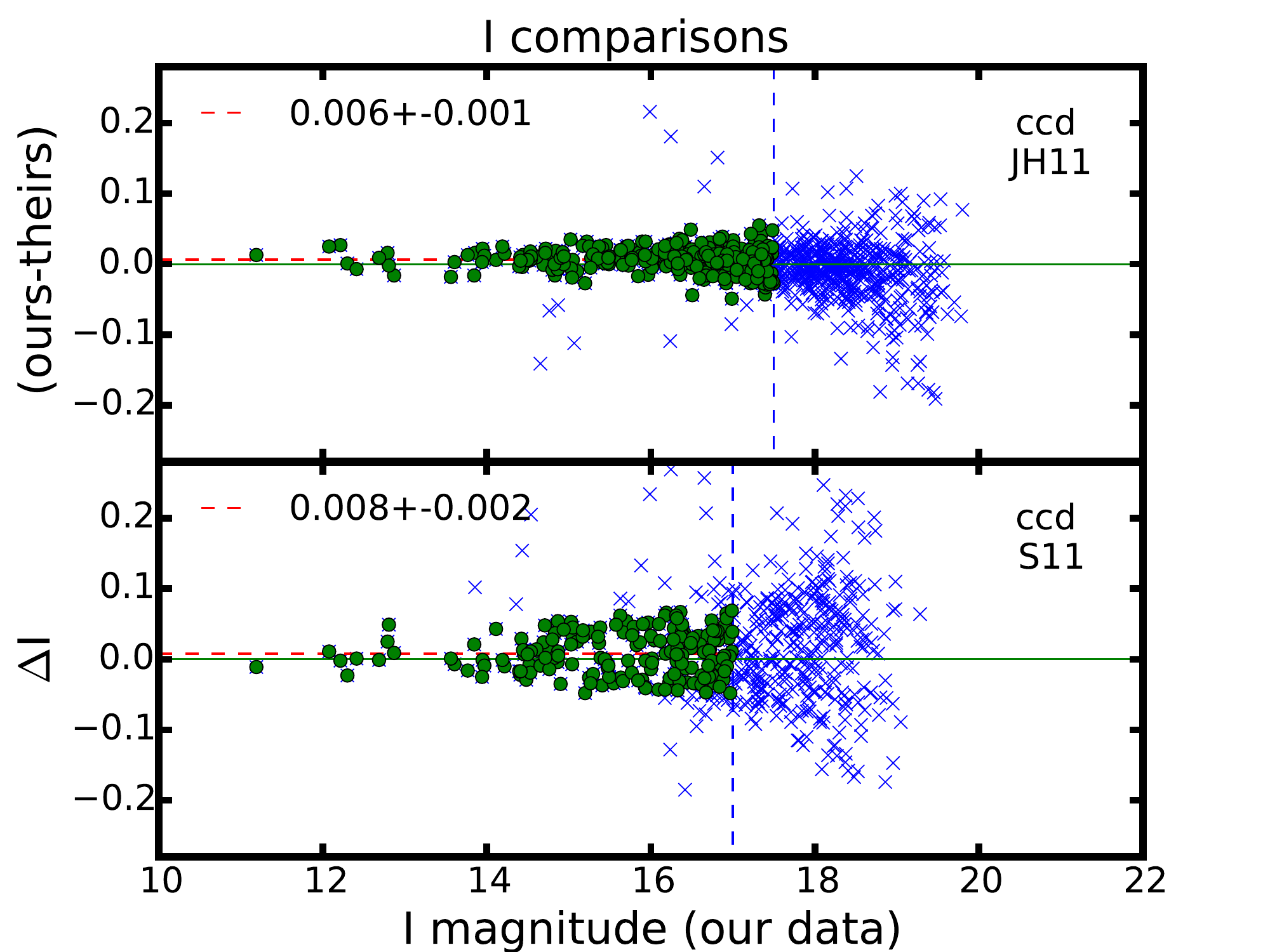}\\
	\caption{Comparison to previous studies in $I$ filters}
	\label{fig:compare I}
\end{figure}

In this section, we compare our photometry to all previous studies that employed the same filters or a subset. We also compare to the Vilnius $V$ of S14, even though the Vilnius $V$ covers a much smaller wavelength range. Figures \ref{fig:compare U} - \ref{fig:compare I} show differences between ours and theirs versus our magnitude. Blue Xs show all matches, while green dots show the subset of generally higher-quality differences that were fit; the faint boundaries were chosen by eye. Dashed red lines show the mean difference $\mu$, also indicated by the number shown, and the error is $\sigma_\mu$. In almost all cases, there is no obvious trend with magnitude, with the possible exception of a few photographic studies. In 16 comparisons, the offsets are less than 0.02 mag; of these, 11 are in excellent agreement with us, having offsets less than 0.01 mag (pe: $B$ from H61, $UBV$ from VH70, $BV$ from JH75; pg: $B$ from S69 and $V$ from VH70; CCD: $V$ from CT91, and $VI$ from JH11). The remaining 13 comparisons have larger offsets, though never larger than 0.063 mag for pe data or 0.017 for CCD data, except for C13. Our offsets with the $UB$ data of C13 are quite large (0.346 mag in $U$ and 0.415 mag in $B$, though smaller, $\sim$ 0.07 mag in $U-B$). The following comparisons are also noteworthy in having little scatter: $BV$ in VH70 and E68 (pe); $B$ in CT91; $BI$ in JH11 and $V$ in S14; and especially $V$ in CT91 and JH11. Paradoxically, some pg scales are significantly different from the pe scale in the same study (for example, pe $U$ differs from pg $U$ by 0.1 mag in VH70).

\section{Analysis and Discussion}  \label{analysis}

In this section, we use photometry of stars in the central region of NGC 7142 (within 4$\arcmin$) to derive cluster parameters. First, using our cluster CMDs, we select a single-star fiducial from the apparent left edge of the main sequence (MS), guided also by Stetson's data for M67 \footnote{http://www.cadc-ccda.hia-iha.nrc-cnrc.gc.ca/en/community/STETSON/standards, data retrieved on 2016 August 3} (2000, hereafter PBS). We then use this fiducial to derive the reddening and metallicity from color-color diagrams. Finally, we determine the age and distance of NGC 7142 by fitting $Y^2$ Isochrones (with Lejeune color calibration) to the MS fiducial in the CMDs. We also carry out a similar and independent analysis of the PBS data for M67, and perform a differential comparison of the two clusters.

\subsection{Determination of the MS Single-star Fiducial Sequence} \label{CMD}
 
\begin{figure}
	
	\centering
	\includegraphics[trim={0.35cm 6.0cm 0cm 0.2cm},clip,width=0.5\textwidth]{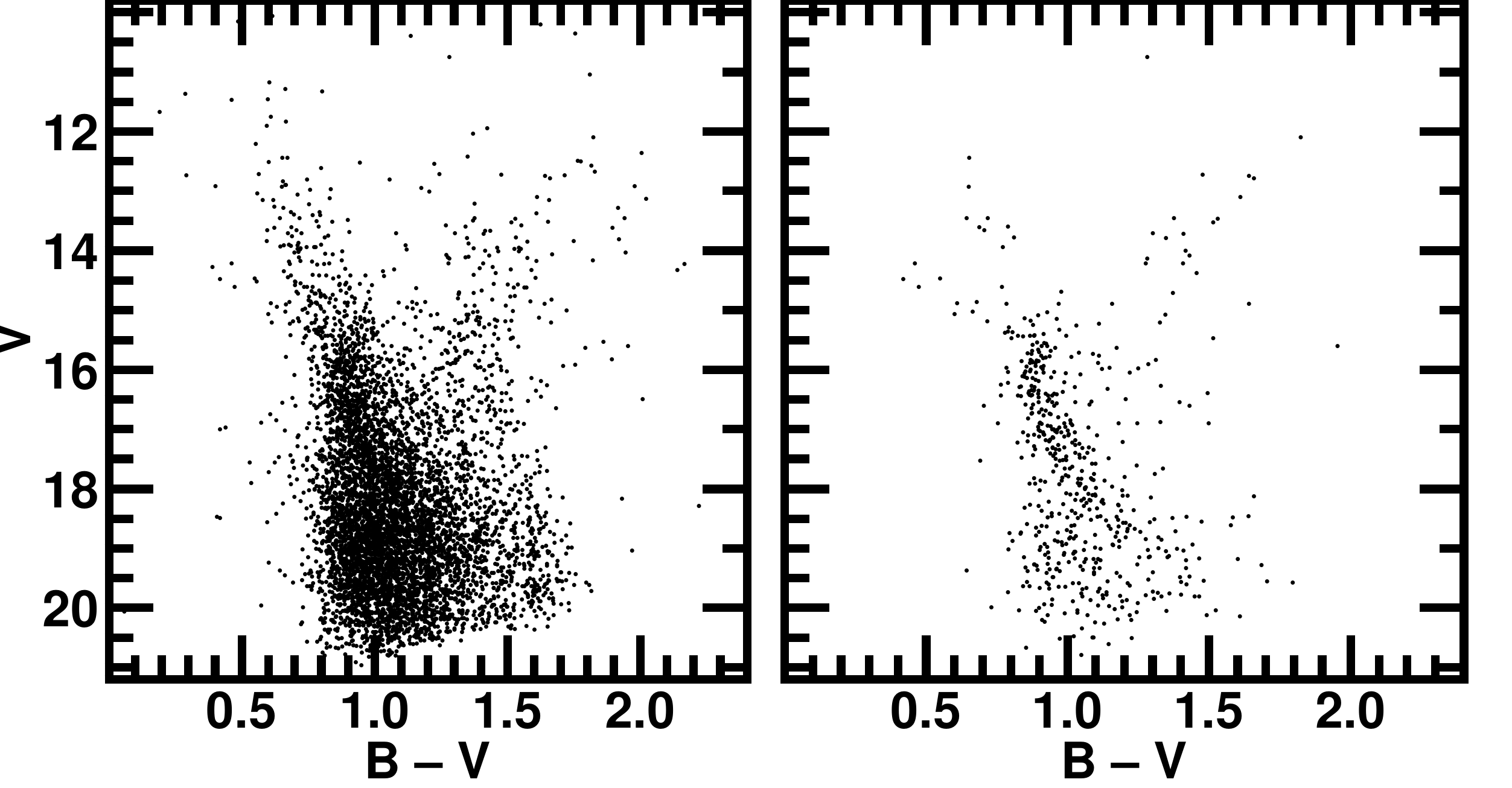} 
	\caption{CMDs for the full HDI field (left) compared to the central field (r $\leq4\arcmin$, right). For both plots, the $y$ axes are $V$ magnitudes and the $x$ axes are $B-V$ colors. Each star is presented as a black dot. The two plots share $x$ and $y$ axes.}
	\label{fig:CMDall_vs_rLT4am}
\end{figure}

Figure \ref{fig:CMDall_vs_rLT4am} (left panel) shows a $V$ vs. $B-V$ CMD for our full HDI field; there is at best only a hint of the MS of the cluster, as the cluster features become obscured by the numerous nonmembers. However, restricting attention to only the central region (r $\leq4\arcmin$, right panel) reveals more clearly a main sequence, TO, subgiant and giant branches, blue stragglers, and red clump stars.

Choosing a single-star fiducial for a differentially reddened cluster can be challenging. Fortunately, the right panel of Figure \ref{fig:CMDall_vs_rLT4am} suggests that the left edge of the MS is fairly well-defined for r $\leq4\arcmin$, so we decided to try to select a fiducial from this left edge. It is hoped that such a fiducial consists mainly of single stars. By contrast, the region to the right of this fiducial may contain binaries of similar reddening and also stars with larger reddening. There will also be confusion near the red hook ($V$ $\sim$ 15.5 mag, $B-V$ $\sim$ 0.8 - 0.9 mag) if stars intrinsically brighter and bluer that have higher reddening get reddened to within the gap or the red hook itself. The possibility exists that some of the few stars to the left of this well-defined left edge might also be cluster members with less reddening, and that the differential reddening across the entire cluster might be larger than that seen in the central (r $\leq4\arcmin$) region, although Figure \ref{fig:compare U} from S14 may hint that this is not so.

The five ($UBVRI$) bands allow ten color combinations. Initially, we used CMDs in $V$ versus color that employed all ten colors. To minimize the uncertainties discussed before, our choice of NGC 7142 fiducial was also guided by the M67 fiducial because the M67 cluster parameters are well-measured and M67 has a metallicity and an age similar to those of NGC 7142 (for more on these similarities, see Sections \ref{Introduction} and \ref{compare}). We chose to use the PBS $UBVI$ data because they include precision photometry of a large number of stars in the M67 field and the all-important $U$ filter, whereas some recent studies \citep{Sandquist04,Yadav08} do not have $U$ measurements or (e.g. \citet{Montgomery93}) are not as precise as the PBS photometry. However, PBS has few $R$ measurements, so we used only the six CMDS from PBS's $UBVI$ to choose (independently of NGC 7142) a single-star fiducial for M67.

\begin{figure}		
	\centering
	\includegraphics[width=0.5\textwidth]{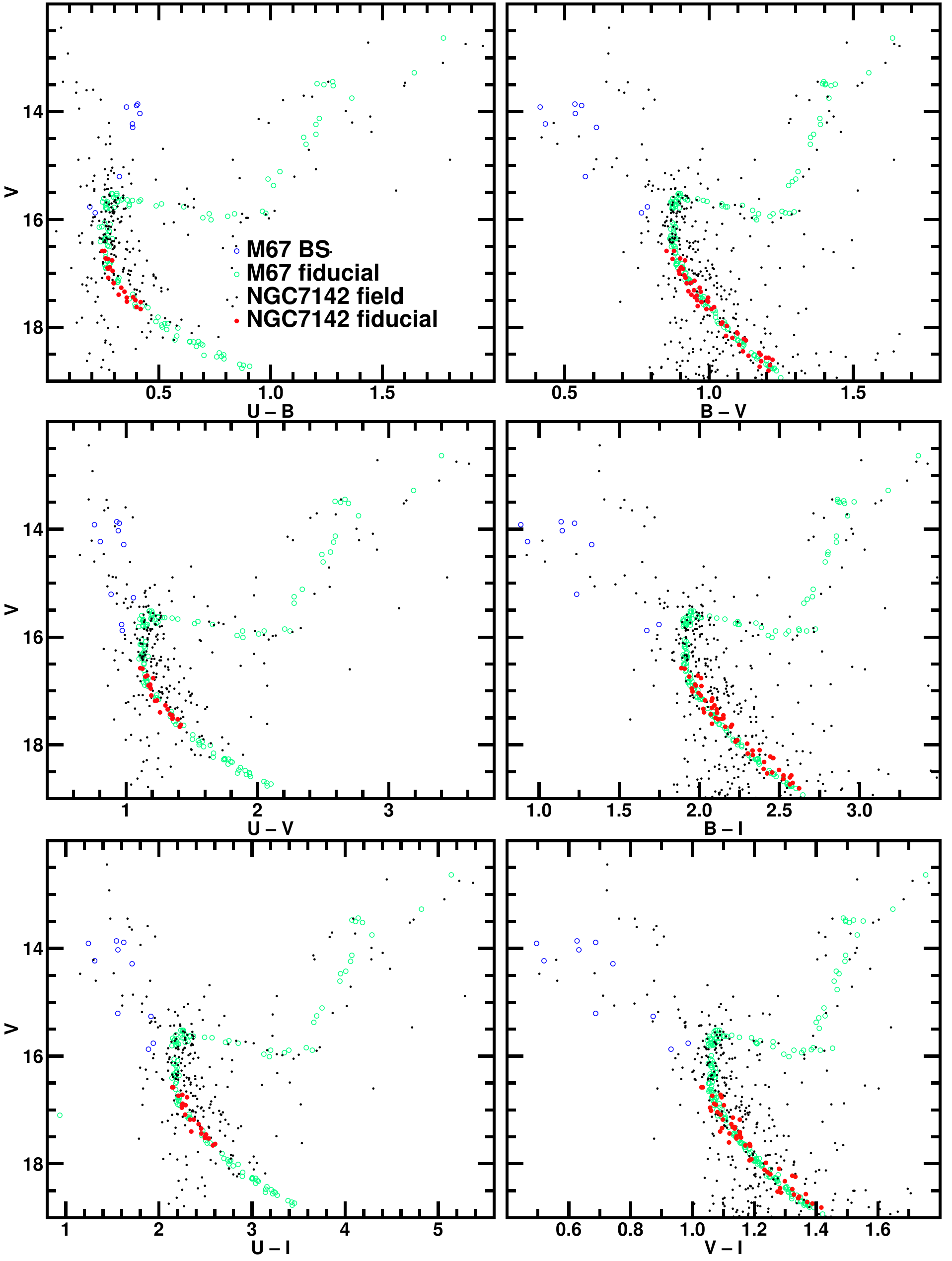} 
	\caption{NGC 7142 single-star MS fiducial selection. M67 fiducial selected from PBS data is shown in green open circles. Red dots are the final NGC 7142 fiducial. Blue open circles are M67 BS stars. Small black dots are stars in the central 4$\arcmin$ of the NGC 7142 field.}
	\label{fig:test_fiducial}
\end{figure}

Figure \ref{fig:test_fiducial} shows our chosen M67 fiducial (green open circles) against the chosen NGC 7142 fiducial (red filled circles, central 4$\arcmin$), and the M67 blue stragglers (blue open circles). The black dots are stars in the central 4$\arcmin$ of the NGC 7142 field. We have eliminated NGC 7142 stars at the top of the main sequence TO because they are evolving off the MS and have surface gravities that are different from those of the unevolved dwarfs of the same $B-V$ color in the Hyades, thereby potentially altering the other color indices. We have also eliminated stars that are so faint that they are difficult to separate from the galactic field. We have kept stars in the range $V$ = 16.6 - 18.7 mag; for $U$ the faint limit is 17.7 mag. We also tried imposing the reddening relations of \citet[C89]{Cardelli1989} by choosing a single value of extra reddening relative to M67 for all CMDs, using the C89-based relations to calculate the expected amount of reddening in each color and then comparing the fiducials. However, not all fiducials then agreed with M67; some were slightly too blue and some were slightly too red. We suspect the problem is not with our photometry, but rather that the relations in C89 may not be entirely accurate at such high values of reddening. Therefore, we allowed the difference in $E(B-V$) between M67 and NGC 7142 to be a free parameter when choosing the fiducial. We then examine all ten colors from $UBVRI$ to eliminate outliers (which may be foreground/background stars or stars with poor photometry) and ended up with the final NGC 7142 fiducial shown in these figures.

\begin{figure*}
	\centering
	\begin{tabular}{cc}
		\includegraphics[width=0.45\textwidth]{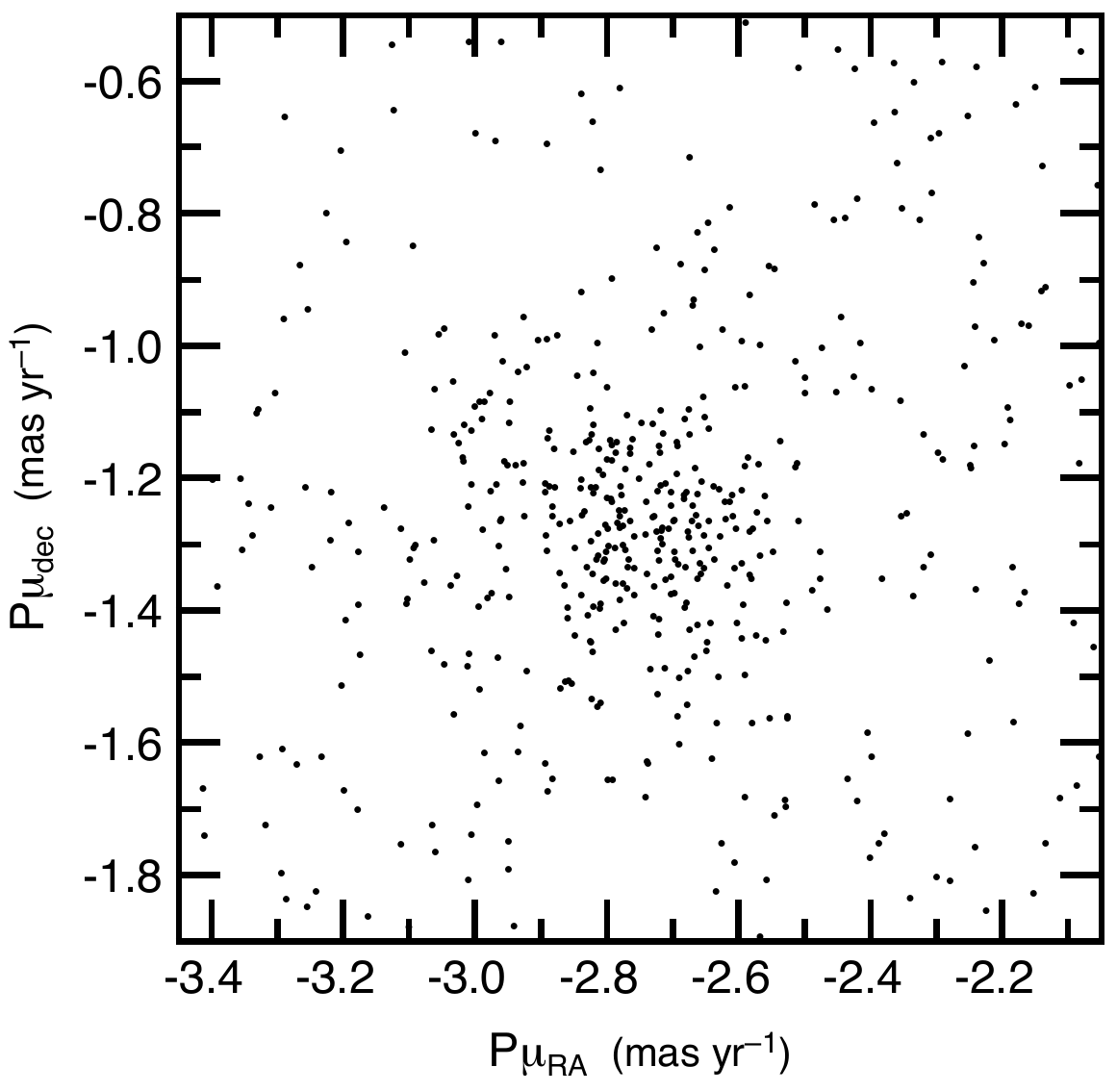}
		\includegraphics[width=0.45\textwidth]{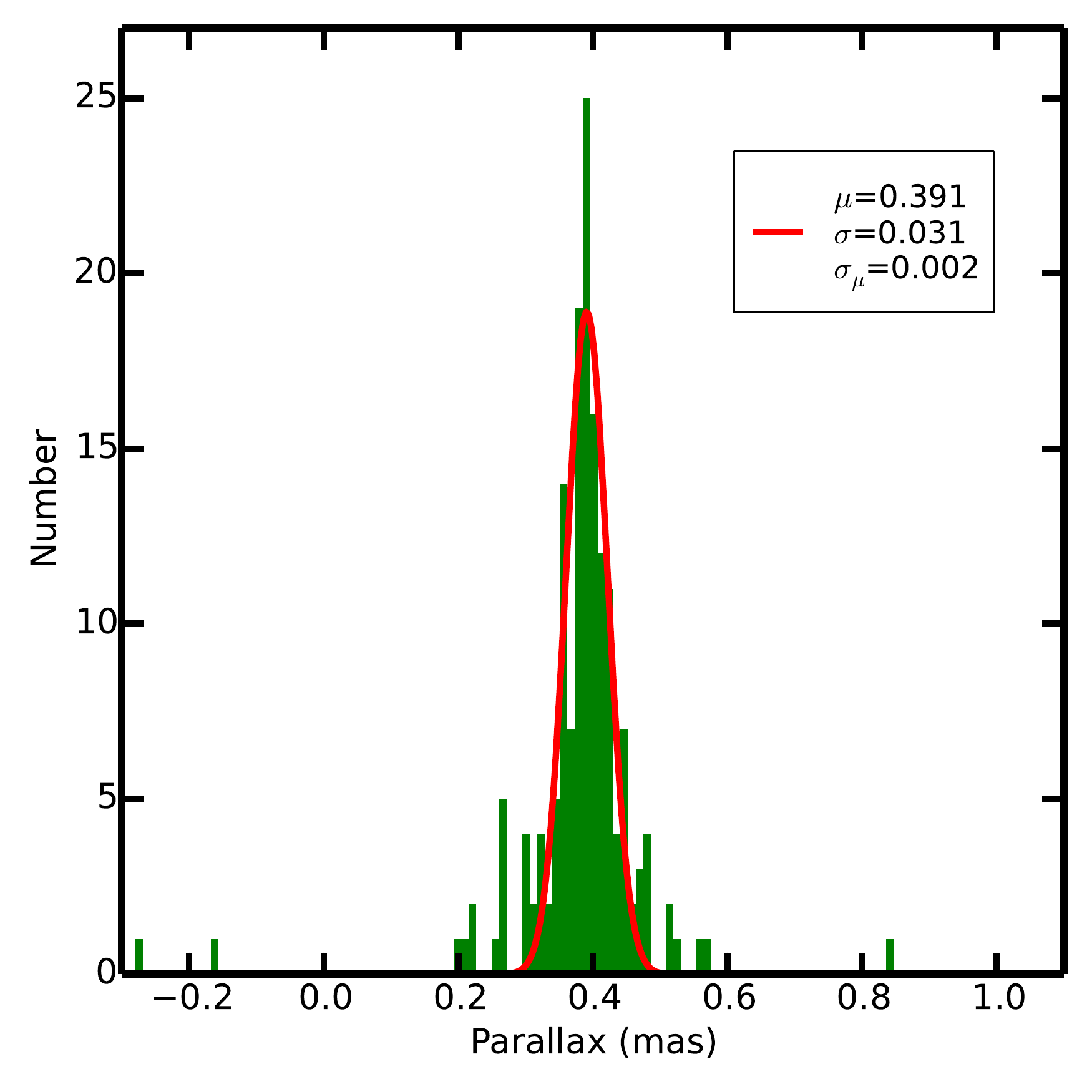}
	\end{tabular}
	\caption{Vector-point diagram using Gaia DR2 proper motion data (left) and the parallax distribution using Gaia DR2 parallaxes (right). Gaussian fit to probable members is shown in red. Derived values for the mean, standard deviation, and standard deviation of the mean are shown in the box.}
	\label{fig:gaia_Pmu}
\end{figure*}

We have used the Gaia DR2 \citep{Gaia16, Gaia18} together with the cluster selections from \citet[CG18]{Cantat18} as a check. This is challenging because our fiducial extends to the faint end of the Gaia data, which has relatively larger errors and is about 0.5 mag fainter (in $G$) than the CG18 sample. We began with a conservative selection (such as shown in Figure \ref{fig:gaia_Pmu}) in which nearly all stars are also clear photometric members on the CMD, but which also missed many photometric members (especially fainter ones, though also some bright ones). We then created a series of new selections while relaxing the selection criteria until the fraction of photometric nonmembers, defined as ones below the apparent main sequence, increased noticeably. These experiments verified that we have succeeded in isolating the left-edge, least reddened portion of the cluster main sequence to use as our fiducial. On the brighter portion of the fiducial, $V\ <$ 17.3 mag, which is the portion used in the color-color diagrams below, nearly all stars are members and only a few of the fainter fiducial stars are not verified as members. This procedure might also be of value in identifying fainter members that have large Gaia errors or are fainter than the Gaia sample. Our fiducial also compares very favorably with the CG18 selection and goes fainter than CG18.

\subsection{Reddening and Metallicity} \label{red}
 
\begin{figure*}
	\centering
	\includegraphics[width=1.0\textwidth]{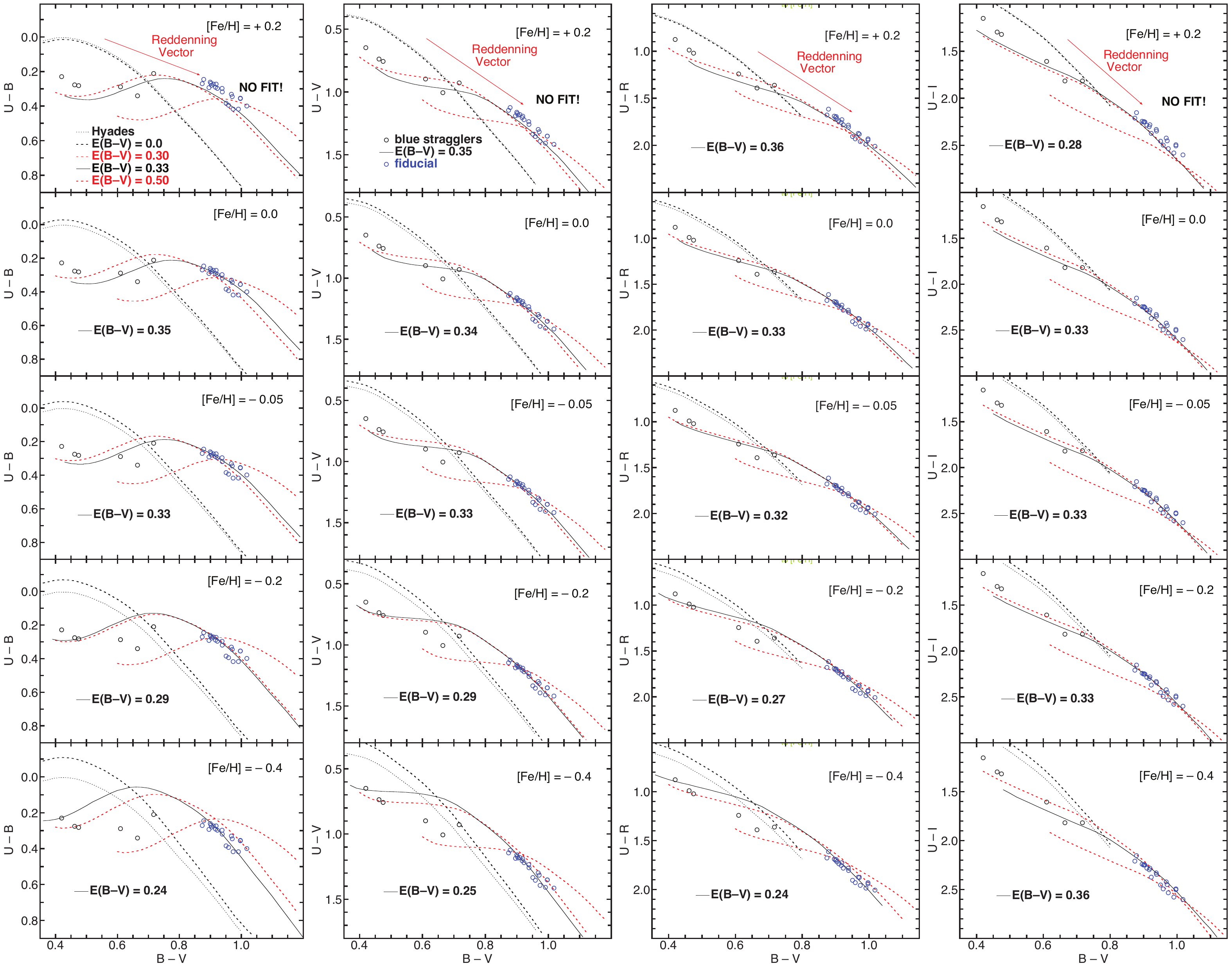} 
	\caption{Reddening and metallicity determination using the $U-B$, $U-V$, $U-R$, and $U-I$ vs. $B-V$ color-color diagrams. Each row represents a fixed metallicity ([Fe/H] = +0.2, 0.0, -0.05, -0.2, -0.4 dex from top to bottom), and each column presents plots using the same color combination. Each column shares the same $x$-axis. As labeled, the Hyades fiducial with [Fe/H] = +0.15 dex is shown as a black dotted line; the zero-reddening curve with the given metallicity is shown as a black dashed line; $E(B-V$) = 0.30 mag and 0.50 mag are shown as red dashed lines; and the best-fit reddening is shown as a black solid line. Blue circles are fiducial members of NGC 7142, and black circles are blue straggler members of NGC 7142.}
	\label{fig:reddening and metallicity}
\end{figure*}

Using the chosen NGC 7142 single-star fiducial sequence from \ref{CMD}, we employed color-color diagrams to estimate the cluster reddening and metallicity simultaneously, following the procedures in \citet[in preparation]{Deli20}. Figure \ref{fig:reddening and metallicity} shows all four colors involving $U$ ($U-B$, $U-V$, $U-R$, $U-I$) plotted against $B-V$; we used these colors because they have larger sensitivity to reddening than do non-$U$ colors. Open black circles represent our MS fiducial for NGC 7142, and blue stragglers are shown as blue open circles. The Hyades fiducial (black dotted line) was used as a zero-reddening reference with assumed [Fe/H] = +0.15 dex \citep{Maderak13, Cummings17}, and metallicity dependence was taken from Yale isochrones (black dashed line). We then calculated reddenings using the relations from C89. It should be noted that errors in C89 may cause errors in the analysis that are difficult to quantify. The red dashed lines are reference reddenings of +0.30 mag and +0.50 mag in $E(B-V$), and the solid black line is our best value for the cluster in a given diagram. These best values were determined as self-consistently as possible from diagram to diagram, by having about the same number of stars above the solid black line as below it. A by-eye estimate of the fitting error is 0.01 - 0.02 mag; for example, in the $U-B$ vs. $B-V$ diagram for [Fe/H] = 0.0 dex, the fit to the fiducial stars for $E(B-V)=0.30$ mag is quite a bit worse than that for $E(B-V)=0.35$ mag. 

Figure \ref{fig:reddening and metallicity} shows that the cluster is too old for both reddening and metallicity to be detemined independently, as the topmost point of the MS are redder than the downturn of the reddening curves. Nevertheless, determining reddenings for several metallicities yields results. We chose to look at [Fe/H] = +0.2, 0.0, -0.2, and -0.4 dex. In each case, we tried to choose the best fit as self-consistently as possible with respect to all the other fits.

In spite of the reddening-metallicity degeneracy, we can rule out high metallicities. For [Fe/H] = +0.2 dex, all the reddening curves fall too high, so there is no reddening value that fits the cluster. This is true for $U-B$, $U-R$, and $U-I$ vs. $B-V$, although $U-V$ is not quite as bad. Furthermore, as we go to lower metallicities, the shape of the reddening curves gets progressively worse with respect to the shape of the NGC 7142 fiducial. These trends argue against considering either higher or lower metallicities than those shown in Figure \ref{fig:reddening and metallicity}. Overall, our best fits are near solar metallicity.

Figure \ref{fig:reddening and metallicity} also shows stars verified by DR2 as blue straggler members of NGC 7142. For $U-B$, $U-V$, and $U-R$, but not $U-I$, versus $B-V$, a similar pattern emerges. For [Fe/H] = +0.2 dex, the blue stragglers are above our chosen best-fit reddening (solid black line), then they are closer for [Fe/H] = 0.0 dex, then below for [Fe/H] = -0.2 dex, and even more below for [Fe/H] = -0.4 dex. This suggests a possible good fit near [Fe/H] = -0.1 dex or perhaps [Fe/H] = -0.05 dex. Therefore, we also show color-color diagrams for [Fe/H] = -0.05 dex (middle panels of Figure \ref{fig:reddening and metallicity}), which may be the best fit among the five metallicities.

Table \ref{tab:red} shows our best-fit value from each diagram. It also shows the average and standard deviation for each metallicity ($\sigma_\mu$). The reddening is just a guess for [Fe/H] = +0.2 dex, because three of the four diagrams cannot produce any kind of fit. Overall, at each metallicity, there is good consistency in the derived reddening values from the various color-color diagrams--and conversely, the largest source of uncertainty in the reddening is due to the uncertainty in the metallicity. Considering $E(B-V$) versus [Fe/H] and ignoring [Fe/H] = +0.2 dex leads to the following relation: 
\begin{align} \label{eqn:EBV}
E(B-V) &=& 0.3375 + 0.3063[Fe/H] + 0.1771[Fe/H]^2
\end{align}
valid only in the range -0.40 dex $\leq$ [Fe/H] $\leq$ +0.10 dex, and assuming the C89 relations are sufficiently valid.

\begin{table*}
	\caption{Results for Reddening (mag) and Metallicity (dex) of NGC 7142}
	\label{tab:red}
	\centering
	\begin{threeparttable}
	\begin{tabular}{|c|c|c|c|c|}
		\hline
		\diagbox[width=8em,trim=l]{Color}{[Fe/H]} & +0.2 & +0.0 & -0.2 & -0.4 \\
		\hline
		$U-B$ & no fit & 0.35 & 0.29 & 0.24\\
		\hline
		$U-V$ & no fit & 0.34 & 0.29 & 0.25\\
		\hline
		$U-R$ & 0.36 & 0.33 & 0.27 & 0.24\\
		\hline
		$U-I$ & no fit & 0.33 & 0.33 \tnote{1}& 0.36\tnote{1}\\
		\hline
		$\mu \pm \sigma$ & no fit & $0.338\pm0.010$ & 0.283$\pm$0.012 & 0.243$\pm$0.006\\
		\hline
	\end{tabular}
	\begin{tablenotes}
		\item [1] Outliers, not used for calculating the average and $\sigma_\mu$.
	\end{tablenotes}
	\end{threeparttable}
\end{table*}

\subsection{Distance and Age} \label{age and distance}
 
\begin{figure*}
	\centering
	\includegraphics[trim={0cm 1.2cm 0cm 0.5cm},clip,width=1.0\textwidth]{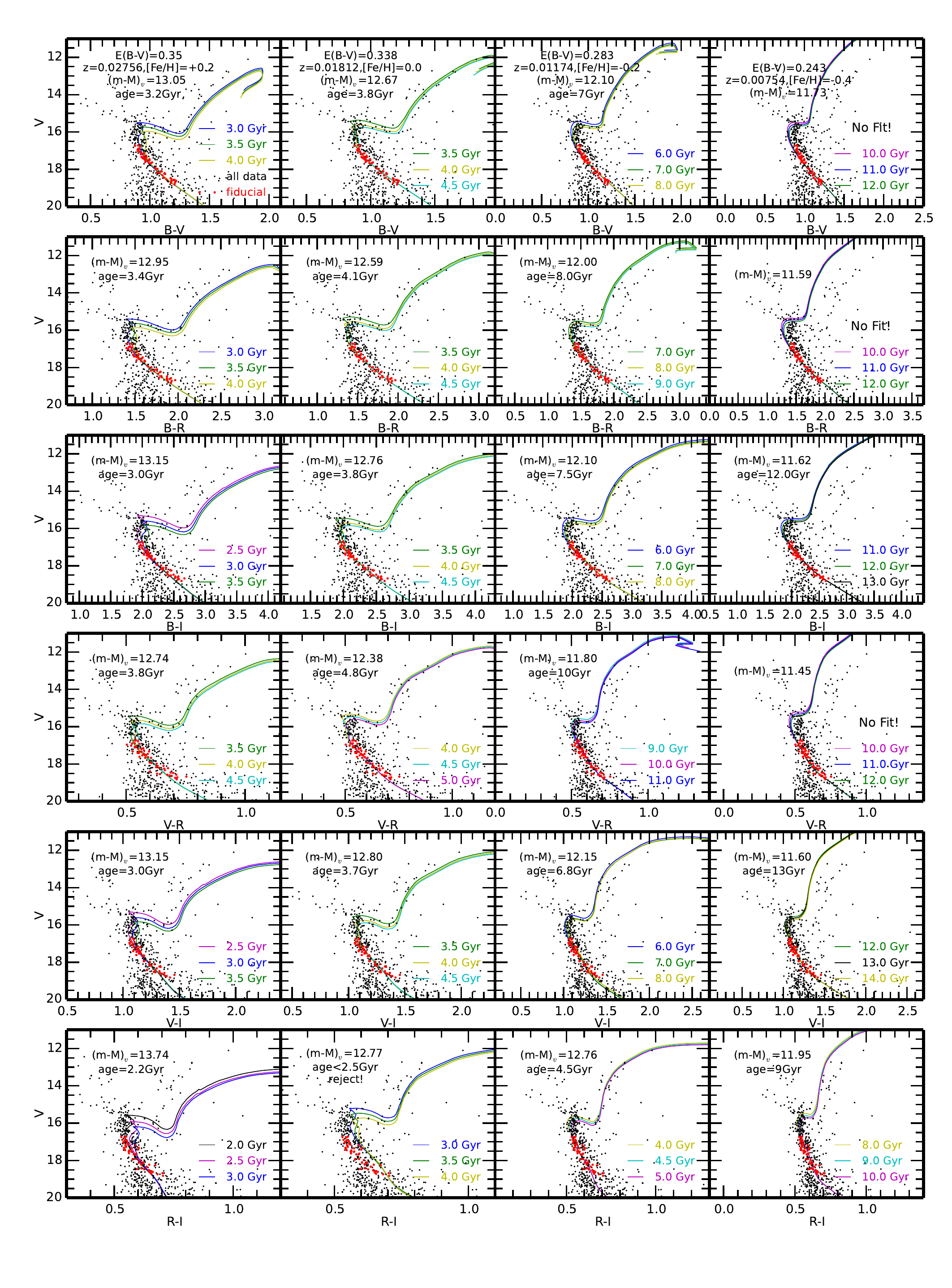}
	\caption{CMDs with $Y^2$ isochrones (Lejeune color calibration). Each column of panels shows a particular metallicity and corresponding reddening (indicated in the top panel). For each panel, values for the best fits for $m-M$ and age are indicated in black. Each isochrone age is given a unique color throughout all panels.}
	\label{fig:Yale Lejeune}
\end{figure*}

For each of the [Fe/H] = +0.2, 0.0, -0.2, and -0.4 dex, we took the corresponding reddening from Table \ref{tab:red} and plotted CMDs using our $BVRI$ photometry to derive the distance and age of NGC 7142. Colors including $U$ were excluded because $Y^2$ isochrones do not fit these as well. Figure \ref{fig:Yale Lejeune} shows these six CMDs for each [Fe/H] -- $E(B-V$) pair: $B-V$, $B-R$, $B-I$, $V-R$, $V-I$ and $R-I$ vs $V$, or 24 plots altogether. The $Y^2$ isochrones were constructed to fit to the data using the Lejeune color calibration, which fits M67 rather nicely; the Yale Green color calibration does not fit the M67 data quite as well. Each plot shows exactly three isochrones, and the age of a given isochrone is indicated in the legend. In each plot, we adjusted the distance modulus ($m-M$) to obtain the best fit to the MS below the TO ($V$ = 17.5 - 18.5 mag); we did so as self-consistently as possible for all 24 plots and then read off the age in each plot. In doing so, we considered how well the best isochrone followed the shape of the red hook ($V$ = 15.5 - 16.5 mag), and also the presumed apparent subgiant members near $V$ = 15.7 mag near the TO to $V$ = 16.0 mag near the base of the red giant branch (RGB). The fact that differential reddening has the potential to lead us astray is also taken into account. So, for example, it is not worrisome that the stars at the base of the RGB near $V$ = 15.9 mag and $B-V$ = 1.3 mag do not match the isochrones well; in fact, these stars are redder than the presumed RGB as well, and it is possible they have slightly larger reddening than the other giants. A by-eye estimate of the fitting errors indicates that they are of order 0.1 mag in distance modulus, and of an age less than the difference between the best-fitting age and the ages of the adjacent isochrones.

Table \ref{tab:Yale Lejeune} shows the results. For consistency, we ended up using only the following four CMDs to get the average distance and age (and their $\sigma_{\mu}$): $B-V$, $B-R$, $B-I$, and $V-I$.

\begin{table*}
	\caption{Results for Distance and Age of NGC 7142}	
	\label{tab:Yale Lejeune}
	\centering
\begin{threeparttable}
		\begin{tabular}{|c|p{1.0cm}p{1.0cm}|p{1.0cm}p{1.0cm}|p{1.0cm}p{1.0cm}|p{1.0cm}p{1.0cm}|}
			\hline
			\multirow{2}{3cm}{Metallicity (dex)   Reddening (mag)} & \multirow{2}{2cm}{\scriptsize [Fe/H] = +0.2  $E(B-V)=0.35$} & & \multirow{2}{3cm}{\scriptsize [Fe/H] = 0.0  $E(B-V)=0.338$} & &\multirow{2}{3cm}{\scriptsize [Fe/H] = -0.2 $E(B-V)=0.283$} & & \multirow{2}{3cm}{\scriptsize [Fe/H] = -0.4 $E(B-V)=0.243$} & \\
			\hline
			\diagbox[width=8em, trim=l]{Color}{(m-M) and Age} & m - M & age  & m - M & age  & m - M & age  & m - M & age\\
			\hline
			$B-V$ & 13.05 & 3.2 $^{3,a}$ & 12.67 & 3.8 $^1$ & 12.10 & 7 $^4$ & ---$^b$ & ---\\
			\hline
			$B-R$ & 12.95 & 3.4 $^3$ & 12.59 & 4.1 $^1$ & 12.00 & 8 $^2$ & --- & ---\\
			\hline
			$B-I$ & 13.15 & 3.0 $^3$ & 12.76 & 3.8 $^1$ & 12.10 & 7.5 $^1$ & 11.62 & 12 $^4$\\
			\hline
			$V-R$ & 12.74 & 3.8 $^2$ $\times$ & 12.38 & 4.8$^4\times^c$ & 11.80 & 10 $^2\times$ & --- & ---\\
			\hline
			$V-I$ & 13.15 & 3.0 $^3$ & 12.80 & 3.7 $^2$ & 12.15 & 6.8 $^2$ & 11.60 & 13 $^4$\\
			\hline
			$R-I$ & 13.74 & 2.2 $^4$ $\times$ & --- & ---  & 12.76 & 4.5 $^1\times$ & 11.95 & 9 $^1\times$\\
			\hline
			$(m - M)$ & $13.08\ \pm$ & $0.09\ ^d$& $12.71\ \pm$ & 0.09 & $12.09\ \pm$ & 0.06 & $11.61\ \pm$ & 0.01 \\
			\hline
			Age(Gyr) & $3.15\ \ \ \pm$ & 0.19 & $3.85\ \ \ \pm$ & 0.17 & $7.32\ \ \ \pm$ & 0.54 & $12.5\ \ \ \pm$ & 0.70 \\
			\hline
			$(m - M)_0$ & 11.99$\pm$ & 0.10 & 11.66$\pm$ & 0.10 & 11.21$\pm$ & 0.07 & 10.86$\pm$ & 0.02 \\
			D(pc) & 2500 $\pm$ & 115 & 2140 $\pm$ & 99 & 1740 $\pm$ & 56 & 1480 $\pm$ & 14 \\
			\hline
		\end{tabular}
		\begin{tablenotes}
			\item[a] Superscript of 1 denotes a good fit, 2 is almost good, 3 is almost poor, and 4 is poor.
			\item[b] ``---"means impossible to fit and/or absurdly large age.
			\item[c] $\times$ indicates outlier and/or poor fit, and was not used in the average.
			\item[d] $(m-M)$ and Age (Gyr) errors are standard deviations; $(m-M)_0$ and D (pc) errors are propagated errors.
		\end{tablenotes}
	\end{threeparttable}
\end{table*}

Not surprisingly, the distance modulus ($m-M$) and ages depend rather sensitively on [Fe/H] (and the corresponding $E(B-V$)). The following quadratics fit these dependences rather well:
\begin{eqnarray}
(m-M) &=& 12.648 + 2.379[Fe/H] - 0.656[Fe/H]^2\\
age(Gyr) &=& 4.01 - 10.169[Fe/H] + 27.969[Fe/H]^2 \label{eqn:age_rela}
\end{eqnarray}

Overall, once again, the fits are best for solar metallicity, which has three ``1" quality factors and a ``2." (But also a ``4" and an unacceptable.) The fits to [Fe/H] of +0.2 and -0.2 dex are not as good, on average, and the ones to [Fe/H] = -0.4 dex can be safely ruled as unacceptable. The ages for [Fe/H] = -0.4 dex are also absurdly old. For giant branches, the isochrones are too red for [Fe/H] = +0.2 dex (especially $R-I$), fairly good for [Fe/H] = 0.0 dex (except $R-I$), too blue for [Fe/H] = -0.2 dex (except $R-I$), and much too blue for [Fe/H] = -0.4 dex.
These observations suggest:
\begin{enumerate}[(1)]
\item solar metallicity is preferred (perhaps to within about +0.1 dex), as are the corresponding reddening, distance, and age;
\item \ [Fe/H] = +0.2 and -0.2 dex are less favored, but perhaps cannot be ruled out altogether, given the various uncertainties;
\item \ [Fe/H] = -0.4 dex can be ruled out fairly safely.
\end{enumerate}

\noindent It follows that:

[Fe/H] = 0.0 $\pm$ 0.1 dex\\
\indent $E(B-V$) = 0.338 $\pm$ 0.031 mag (fiducial, left edge of \indent MS)\\
\indent $m-M$    = 12.65 $\pm$ 0.23 mag\\
\indent age    = $4.0^{-0.7}_{+1.3}$ Gyr\\
\indent $(m - M)_o$ = 11.60 $\pm$ 0.25 mag\\
\indent $D$ = 2090 $\pm$ 240 pc.\\

The error estimates for $E(B-V$), $m-M$, and age are those propagated from an error of 0.1 dex in [Fe/H], which dominates over the internal errors listed in Tables \ref{tab:red} and \ref{tab:Yale Lejeune}. The error estimates for $(m-M)_o$ and D also include the propagated error in $E(B-V$).

Various uncertainties may affect the shape of the isochrones--and thus the quality of the fits to the data, such as uncertainties in the color calibration, among others. However, we find the isochrones do fit M67 fairly well (section \ref{compare}). That is a cluster whose parameters are known much better than those of NGC 7142, suggesting that the above conclusions are valid at least relative to M67.

\subsection{Comparison to Previous Studies}

In comparing to previous studies (see Section \ref{Introduction}), recall that none of the previous studies employed MS stars that are clearly below the MS TO, and a number of the photometric studies did not use an ultraviolet filter, which can be sensitive to reddening and metallicity. Previous studies encompass the following ranges for the cluster parameters:

$E(B-V$) = 0.18 - 0.51 mag \\
\indent $ \lbrack Fe/H \rbrack$ = -0.45 - +0.14 dex \\
\indent $m-M$ = 11.8 - 13.7 mag \\
\indent ($m-M$)$_o$ = 10.9 - 12.5 mag \\
\indent age = 2 - 7 Gyr. \\

Overall, our results fall in the middle of these ranges. For reddening, we agree well with the CT91 reanalysis of some of the JH75 data ($E(B-V$) = 0.35 mag), as well as JH11 ($E(B-V$) = 0.32 $\pm$ 0.05 mag) and S14 ($E(B-V$) $\sim$ 0.35 mag); we agree somewhat less well with JH75 ($E(B-V$) = 0.25 mag) and S11 ($E(B-V$) = 0.25 $\pm$ 0.06 mag); and we disagree with the other studies. Our [Fe/H] is consistent with the range from \citet{Friel02}(-0.10 $\pm$ 0.10 dex, a recalibration of the FJ93 study) to \citet{Twarog97}(+0.04 $\pm$ 0.06 dex) to the high-resolution spectroscopic study of J07 (+0.08 $\pm$ 0.06 dex), but is less consistent with the other studies. Our age is compatible with that of \citet{Carraro98}(4.9 Gyr), \citet{Salaris04}(4 $\pm$ 1 Gyr), and S14 (3.0 $\pm$ 0.5 Gyr), but less so with JH11 (6.9 Gyr). We also find that NGC 7142's metallicity and age are indistinguishable from those of M67 (section \ref{compare}), in contrast to those who stated evidence that NGC 7142 is younger than M67 (S11) or older than M67 (CT91).

\subsection{An Independent Analysis of M67 and Comparisons with NGC 7142} \label{compare}

A number of high-resolution spectroscopic studies of M67 have found solar [Fe/H] for that cluster; for example, \citet{Garcia98} found +0.04 $\pm$ 0.04, \citet{Hobbs91} found -0.04 $\pm$ 0.12, \citet{Friel92} found +0.02 $\pm$ 0.12, \citet{Santos09} found +0.02 $\pm$ 0.01 for giants and +0.01 $\pm$ 0.04 for dwarfs, and \citet{Jacobson11} found -0.01 $\pm$ 0.05. In addition, studies also using $Y^2$ isochrones or closely related models have found ages near 4 Gyr; for example, \citet{Dinescu95}, \citet{Sarajedini99} and \citet{Sills00} all find 4.0 $\pm$ 0.5 Gyr. These values for the M67 metallicity and age are indeed quite similar to what we have found for NGC 7142. It is thus of interest to perform both an independent re-evaluation of M67's parameters using our methods, as well as a more detailed differential comparison between M67 and NGC 7142.

To determine the reddening of M67, we begin with a color-color analysis similar to that for NGC 7142, using the PBS $UBVI$ data (PBS had very few $R$ data). Using the three color-color diagrams involving $U$, we find [Fe/H] = -0.02 $\pm$ 0.05 dex and $E(B-V$) = 0.04 $\pm$ 0.01 mag. The metallicity is consistent with those listed above, and the reddening is consistent with the value $E(B-V$) = 0.041 $\pm$ 0.004 mag from \citet{Taylor07}. \footnote{Note that, if we use the MMJ $UVBI$ data, we find, overall, [Fe/H] $\sim$ 0.0 dex to within about 0.01 dex and $E(B-V$) = 0.05 mag to within about 0.01 mag. These are also consistent with the studies mentioned above, though there is more scatter than when using the PBS data. One could argue that the $U-B$ diagram suggests that a slightly subsolar [Fe/H] (and slightly lower reddening) might yield a better fit, while the $U-I$ diagram implies a slightly supersolar [Fe/H] (and slightly higher reddening). The MMJ data for the blue stragglers do give a similar reddening, whereas the PBS data for those stars are (mostly) consistent with the value of $E(B-V$) = 0.04 mag found above.} We refer the reader to \citet{Taylor07} for a very thorough review of many reddening studies prior to 2007. To arrive at his result, \citet{Taylor07} performed a detailed analysis of the reddening of M67 using data from a variety of methods. Since our result is consistent with his, we will henceforth simply adopt his value of $E(B-V$) = 0.041 $\pm$ 0.004 mag. Fits of $Y^2$ isochrones to the three CMDs ($V$ versus $B-V$, $B-I$, and $V-I$, PBS data) for M67 then yield $m-M$ = 9.75 $\pm$ 0.03 mag, and age$_{\rm{M67}}$ = 3.85 $\pm$ 0.17 Gyr. If we use only the same three CMDs to get the age of NGC 7142 we find, age = 3.778 $\pm$ 0.064 Gyr.

Our analyses for the two clusters show that their metallicities are indistinguishable and near-solar, although the uncertainty for the [Fe/H] of NGC 7142 is arguably larger than that for M67. The ages of the two clusters are also indistinguishable, assuming that they have the same metallicity. However, if the clusters have different metallicity, then they may also have different ages. For example, equation (\ref{eqn:age_rela}) gives the following ages for NGC 7142, for the indicated metallicities:
\begin{align}
[Fe/H] = +0.1\ dex,   age = 3.27\ Gyr \nonumber \\
[Fe/H] =  0.0\ dex,   age = 4.01\ Gyr \nonumber\\
[Fe/H] = -0.1\ dex,   age = 5.27\ Gyr \nonumber
\end{align}

For our differential comparison of the two clusters, we use the same six CMDs as in section \ref{CMD} (Figure \ref{fig:test_fiducial}; $V$ versus the six colors that can be formed from $UBVI$ data). In each CMD, the fiducial of M67 was shifted in $V$ and in color (as self-consistently as possible) until it matched the fiducial of NGC 7142. The fiducials agree very well, though recall that this was partly by design (Section \ref{CMD}). Figure \ref{fig:m67_PBS_comp} shows the three non-$U$ CMDs ($V$ versus $B-V$, $B-I$, and $V-I$) together with solar-metallicity $Y^2$ isochrones. Several things are apparent:

\begin{figure*}
	\centering
	\includegraphics[clip,width=1.\textwidth]{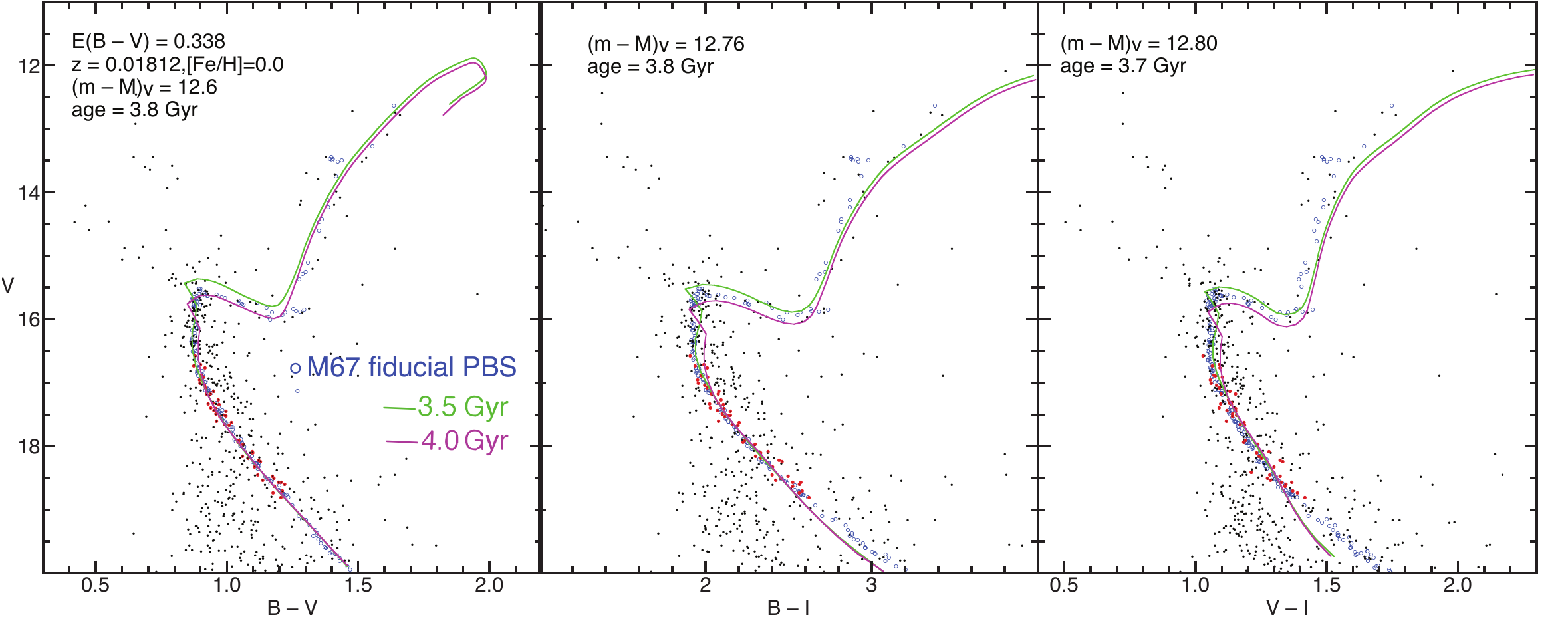}
	\caption{Differential Comparisons of M67 and NGC 7142 using $BVI$ data}
	\label{fig:m67_PBS_comp}
\end{figure*}

\begin{enumerate}[1.]
\item In the $V$ versus $B-V$, $B-I$, and $V-I$ CMDs, the giant branches agree remarkably well for the brighter stars ($V\ <$ 14 mag). Near $V$ = 14 mag, three stars of NGC 7142 agree well with M67 in $B-V$, but are slightly red in $B-I$ and $V-I$. Fainter than $V$ = 14 mag and down to the base of the giant branch (near $V$ = 16 mag), the NGC 7142 giants are mostly slightly redder than the M67 ones; this might indicate a small amount of differential reddening for these stars.
\item There is an apparent population of subgiants in the NGC 7142 field ($V$ = 15.6 mag and $B-V$ = 0.9 mag to $V$ = 16.0 mag at $B-V$ = 1.2 mag) that corresponds well with the M67 fiducial subgiants. Initially, we made the reasonable assumption that these are mostly subgiant members of NGC 7142 and used these stars to help us with the isochrone fits and determination of ages. Information from the Gaia DR2 has verified that most of these stars are subgiant members (see also CG18).
\item Although the RGBs have good agreement in $B-V$, $B-I$, $V-I$, and perhaps $U-I$, they do not agree quite as well in $U-B$ and $U-V$. This might suggest an issue with the calibration for either our or the PBS $U$ filter for redder stars.
\item Some notes about the alignment of the fiducials: If one pays particular attention to the pattern of M67 subgiants above the CMD gap ($V$ = 15.6 - 15.9 mag), one can easily imagine NGC 7142 following this pattern closely. One can also imagine a CMD gap in NGC 7142 at $V$ = 16.0 mag, similar to the one in M67, obscured slightly by a few stars that are differentially reddened and thus fall in the gap. At first it might be tempting to consider the flat line of stars at $V$ = 15.4 mag going from $B-V$ = 0.8 to 0.95 mag as subgiant members of NGC 7142 just above the CMD gap, which includes a few stars bluer than the blue side of the hook and slightly fainter, as we see in M67. However, similar M67 subgiants do not show constant luminosity. The M67 subgiant luminosity rises as one goes from bluer to redder above the CMD gap (from $V$ = 16.0 mag to 15.5 mag), and the theoretical isochrones also predict this. Furthermore, there are no corresponding NGC 7142 subgiants at this luminosity ($V\ \sim$ 15.5 mag) that are redder, as would be expected for a cluster of this age--and as are seen along a 3.8 Gyr isochrone. Finally, the red hook would then be unusually large. So, with the correspondence between the two clusters as chosen, one might conclude that these brighter stars are blue stragglers (or nonmembers). In fact, information from the Gaia DR2 suggests that most are nonmembers while a few are member blue stragglers.
\item The giant branch and the MS stars with $V$ = 16 - 18 mag suggest, perhaps only mildly, that the differential reddening (in this central region) is no more than roughly 0.1 dex in $E(B-V$).
\end{enumerate}

This differential comparison allows a semi-independent evaluation of the reddening of NGC 7142, assuming a known value for the reddening of M67. In particular, we have taken the shifts in the colors of the M67 fiducial required to match the NGC 7142 fiducial as indicating the excess reddening of NGC 7142 relative to M67, and added those to the reddening of M67. In so doing, we have used the relations in C89 to transform the reddening in the different colors ($E(U-B$), $E(U-V$), $E(U-I$), $E(B-V$), $E(B-I$), $E(V-I$)) into $E(B-V$). This method differs fundamentally from that used in section \ref{red}, which employs color-color diagrams. Another feature of the method is that it adds the PBS photometry to the analysis, rather than relying solely on our own photometric data. Finally, it makes use of the arguably now well-known reddening of M67 \citep{Taylor07}. Table \ref{tab:red_m67} shows the results.\footnote{We repeated this exercise using the MMJ data and found a very similar result--but with larger variations.} The average reddening after ignoring an outlier is $E(B-V$) = 0.339 $\pm$ 0.008 mag, which is remarkably close to the value $E(B-V$) = 0.338 $\pm$ 0.010 mag derived earlier (for solar metallicity) using the color-color diagrams of NGC 7142 in section \ref{red}. If the outlier had been included, we would have found $E(B-V$) = 0.343 $\pm$ 0.008 mag, still in excellent agreement with the value of section \ref{red}.

We now consider our results in the context of recent discussions about zero-point offsets in the Gaia parallaxes. Consistent with the suggestions of \citet{Stassun18}, \citet{Riess18}, \citet{Cantat18}, \citet{Chane19}, and \citet{Zinn19} that the Gaia DR2 parallaxes are too small by about 0.05 - 0.08 mas, \citet{Lindegren18} find mean quasar parallaxes of -0.029 mas (they should be zero). The magnitude of the offset depends on source brightness and color, among other possible variables, and is typically in the range 0.03 - 0.06 mas too small \citep{Leung19}. The comparison of \citet{Arenou18} to a number of external catalogs yields a similar result. Clusters that are spatially very small might have significantly larger offsets, while some effects may average out for large clusters. We are thus curious to compare the parallaxes we infer for our two clusters to their Gaia parallaxes.

For NGC 7142, our distance derived above translates to $\pi = 0.478 \pm 0.060$ mas. From a selection from Gaia DR2 that includes only the most reliable members, particularly stars that satisfy the criteria -2.9 mas yr$^{-1} \leq P\mu_{R.A.} \leq$ -2.6 mas yr$^{-1}$ and -1.4 mas\ yr$^{-1} \leq P\mu_{decl.} \leq$ -1.1 mas yr$^{-1}$ (see Figure \ref{fig:gaia_Pmu}, Left panel), we find the cluster's Gaia parallax to be 0.391 $\pm$ 0.002 mas (see Figure \ref{fig:gaia_Pmu}, right panel). The difference in the sense (us--Gaia) is 87 $\pm$ 60 $\mu$as. For M67, using the \citet{Taylor07} value for reddening (and error) yields $(m-M)_o$ = 9.623 $\pm$ 0.030 mag, $D$ = 841 $\pm$ 11 pc, and $\pi$ = 1.188 $\pm$ 0.016 $\mu$as. The difference in the sense (us--Gaia) is $48 \pm 15$ $\mu$as. Both of these values are consistent with what has been found in the studies discussed above.

\begin{table}
	\caption{Reddening $E(B-V$) (mag) in NGC 7142 using PBS M67 data and C89 results}
	\footnotesize
	\label{tab:red_m67}
	\centering
	\begin{threeparttable}
		\begin{tabular}{|c|c|c|}
		\hline
		\diagbox[width=8em, trim=l]{Color}{Reddening} & $ E(B-V)$ \tnote{1} & $E_0(B-V)$ \tnote{2}\\
		\hline
		$U-B$ & 0.35 \tnote{3} & 0.391 \tnote{3}\\
		\hline
		$U-V$ & 0.31 & 0.351\\
		\hline
		$U-I$ & 0.30 & 0.341\\
		\hline
		$B-V$ & 0.30 & 0.341\\
		\hline
		$B-I$ & 0.29 & 0.331\\
		\hline
		$V-I$ & 0.29 & 0.331\\
		\hline
		$\mu\ \pm\ \sigma$ & 0.298 $\pm$ 0.008 & 0.339 $\pm$ 0.008\\
		\hline
		\end{tabular}
	\begin{tablenotes}
		\item[1] Reddening with respect to PBS M67 reddening assuming C89 factor.
		\item[2] Reddening after adding M67 reddening $E(B-V$) = 0.041 mag \citep{Taylor07} to Column 2.
		\item[3] We throw out this outlier when calculating the average.
    \end{tablenotes}
    \end{threeparttable}
\end{table}

\section{Conclusions} \label{summary}

We present CCD $UBVRI$ photometry of 8702 stars taken with the WIYN 0.9m telescope and HDI in the direction of the old open cluster NGC 7142, covering an approximate area of $0.5 ^\circ \times0.5 ^\circ$. Seventy images were taken at five different exposure levels, resulting in the following magnitude ranges: 10.6 - 20.4 mag in $U$, 10.6 - 22.0 mag in $B$, 10.0 - 21.8 mag in $V$, 9.2 - 20.7 mag in $R$, and 8.5 - 19.9 mag in $I$. Comparison with previous studies finds the best agreement to be with some of the photoelectric and CCD studies. 

Selection of a single-star fiducial sequence is complicated by the cluster's differential reddening. We verified previous suggestions that the well-studied, old open cluster M67 appears photometrically similar to NGC 7142, and used CMDs with as many colors as possible (comprised of $V$ versus $UBVRI$ colors), guided also by Stetson's M67 photometry \citep{Stetson87}, to select a single-star fiducial sequence for the left (minimally reddened) edge of the NGC 7142 main sequence. This fiducial is confirmed by Gaia DR2 data to be a left-edge (lowest reddening) fiducial of the cluster. We use four colors involving $U$, versus $B-V$, to determine the reddening and metallicity of the cluster. Even though the advanced age of the cluster complicates a precise determination of both parameters (the usual kink in these color-color diagrams is missing, so there is some degeneracy between reddening and metallicity), we determine the following reddening-metallicity relation: 
\begin{eqnarray*}
E(B-V) &=& 0.3375 + 0.3063[Fe/H] + 0.1771[Fe/H]^2
\end{eqnarray*}
valid in the range -0.40 $\leq$ [Fe/H] $\leq$ +0.10.

Furthermore, the quality of the fits suggests a preferred range of [Fe/H] = 0.0 $\pm$ 0.1 dex. Consideration of the presumed cluster blue stragglers supports these conclusions.

Comparison of the six CMDs formed from $V$ versus $BVRI$ colors to Yonsei-Yale isochrones \citep{Demarque04} for various metallicity-reddening pairs from the above relation yields the following relations:
\begin{eqnarray*}
(m-M) &=& 12.648 + 2.379[Fe/H] - 0.656[Fe/H]^2\\
age(Gyr) &=& 4.01 - 10.169[Fe/H] + 27.969[Fe/H]^2
\end{eqnarray*}

\noindent The preferred range of [Fe/H] = 0.0 $\pm$ 0.1 dex yields the following ranges:

$E(B-V$) = 0.338 $\pm$ 0.031 mag (fiducial, left edge of \indent MS)\\
\indent $m-M$    = 12.65 $\pm$ 0.23 mag\\
\indent age    = 4.0$^{-0.7}_{+1.3}$ Gyr\\
\indent $(m-M)_o$ = 11.60 $\pm$ 0.25 mag\\
\indent $D$ = 2090 $\pm$ 240 pc,

\noindent where the error estimates for $E(B-V$), $m-M$, and age are those propagated from an error of 0.1 dex in [Fe/H]. The error estimates for $(m-M)_o$ and D also include the propagated error in $E(B-V$).

We have also performed a re-evaluation of the parameters of M67 using our methods and Stetson's $UBVI$ data, as well as a detailed comparison to M67. Using the three color-color diagrams involving $U$, we find [Fe/H] = -0.02 $\pm$ 0.05 dex, consistent with previous high-resolution spectroscopic studies, and $E(B-V$) = 0.04 $\pm$ 0.01 mag, consistent with the value $E(B-V$) = 0.041 $\pm$ 0.004 mag from the thorough, detailed study of \citet{Taylor07}. Comparison of the three CMDs ($V$ versus $B-V$, $B-I$, and $V-I$) to $Y^2$ isochrones \citep{Kim02, Demarque04} then yields $m-M$ = 9.75 $\pm$ 0.03 mag and age$_{\rm{M67}}$ = 3.85 $\pm$ 0.17 Gyr; the three same CMDs for NGC 7142 yield age = 3.778 $\pm$ 0.064 Gyr. We thus find that the metallicity and age of M67 and NGC 7142 are indistinguishable. However, we have discussed how the metallicity of NGC 7142 is arguably more uncertain than that of M67, and if it differs from that of M67, then the age may also differ.

The similarity of the two clusters and the fact that M67 is well-studied allows for a semi-independent evaluation of the parameters of NGC 7142, by assuming those for M67 and conducting a differential analysis between the two clusters. Assuming $E(B-V$) = 0.041 $\pm$ 0.004 mag for M67 and shifting the fiducials in the $V$ versus $U-B$, $U-V$, $U-I$, $B-V$, $B-I$, and $V-I$ CMDs in color (and $V$) until they match yields the excess reddening of (the left edge of) NGC 7142 relative to M67; altogether, for NGC 7142, we find $E(B-V$) = 0.339 $\pm$ 0.008 mag, in superb agreement with the value $E(B-V$) = 0.338 $\pm$ 0.031 mag listed above.

The differences between our inferred parallaxes and the Gaia DR2 values are $87 \pm 60$ $\mu$as for NGC 7142 and 48 $\pm$ 15 $\mu$as for M67, consistent with previous studies.

\section{Acknowledgments}   \label{acknowledgments}

C.P.D. acknowledges support from the NSF through grants AST-1211699 and AST-1909456. B.J.A.T. and B.A.T. acknowledge support from the NSF through grant AST-1211621.

\end{document}